\documentclass{bmcart}

\usepackage[utf8]{inputenc} %unicode support
\usepackage[colorlinks,citecolor=blue,urlcolor=blue,bookmarks=false,hypertexnames=true]{hyperref} 

\usepackage{graphicx}

\startlocaldefs
\endlocaldefs

\begin{document}

%%% Start of article front matter
\begin{frontmatter}

\begin{fmbox}
\dochead{Research}


\title{Digital Landscape of COVID-19 Testing: Challenges and Opportunities}
%%
%% not used. Repeat \author command as much %%
%% as required.                             %%
%%                                          %%
%%%%%%%%%%%%%%%%%%%%%%%%%%%%%%%%%%%%%%%%%%%%%%

\author[
   addressref={pcf},                   % id's of addresses, e.g. {aff1,aff2}
]{\fnm{Darshan} \snm{Gandhi}}
\author[
   addressref={pcf},                   % id's of addresses, e.g. {aff1,aff2}
]{\fnm{Rohan} \snm{Sukumaran}}
\author[
   addressref={pcf},                   % id's of addresses, e.g. {aff1,aff2}
]{\fnm{Priyanshi} \snm{Katiyar}}
\author[
   addressref={pcf, ITGH},    
]{\fnm{Alex} \snm{Radunsky}}
\author[
   addressref={pcf},                   % id's of addresses, e.g. {aff1,aff2}
]{\fnm{Sunaina} \snm{Anand}}
\author[
   addressref={pcf,NIH},                   % id's of addresses, e.g. {aff1,aff2}
]{\fnm{Shailesh} \snm{Advani}}
\author[
   addressref={pcf},                   % id's of addresses, e.g. {aff1,aff2}
]{\fnm{Jil} \snm{Kothari}}
\author[
   addressref={pcf,mit,harvardash},                   % id's of addresses, e.g. {aff1,aff2}
]{\fnm{Kasia} \snm{Jakimowicz}}
\author[
   addressref={pcf},                   % id's of addresses, e.g. {aff1,aff2}
]{\fnm{Sheshank} \snm{Shankar}}
\author[
   addressref={pcf},                   % id's of addresses, e.g. {aff1,aff2}
]{\fnm{Sethuraman} \snm{T. V.}}
\author[
   addressref={pcf},                   % id's of addresses, e.g. {aff1,aff2}
]{\fnm{Krutika} \snm{Misra}}
\author[
   addressref={pcf,UCB},                   % id's of addresses, e.g. {aff1,aff2}
]{\fnm{Aishwarya} \snm{Saxena}}
\author[
   addressref={pcf},                   % id's of addresses, e.g. {aff1,aff2}
]{\fnm{Sanskruti} \snm{Landage}}
\author[
   addressref={pcf},                   % id's of addresses, e.g. {aff1,aff2}
]{\fnm{Richa} \snm{Sonker}}
\author[
   addressref={pcf},                   % id's of addresses, e.g. {aff1,aff2}
]{\fnm{Parth} \snm{Patwa}}
\author[
   addressref={pcf},                   % id's of addresses, e.g. {aff1,aff2}
]{\fnm{Aryan} \snm{Mahindra}}
\author[
   addressref={pcf},                   % id's of addresses, e.g. {aff1,aff2}
]{\fnm{Mikhail} \snm{Dmitrienko}}
\author[
   addressref={pcf},                   % id's of addresses, e.g. {aff1,aff2}
]{\fnm{Kanishka} \snm{Vaish}}
\author[
   addressref={pcf},                   % id's of addresses, e.g. {aff1,aff2}
]{\fnm{Ashley} \snm{Mehra}}
\author[
   addressref={pcf},                   % id's of addresses, e.g. {aff1,aff2}
]{\fnm{Srinidhi} \snm{Murali}}
\author[
   addressref={pcf},                   % id's of addresses, e.g. {aff1,aff2}
]{\fnm{Rohan} \snm{Iyer}}
\author[
   addressref={pcf, stonybrook},              % id of corresponding address, if any
]{\fnm{Joseph} \snm{Bae}}
\author[
   addressref={medialab, mit, harvardmed},                   % id's of addresses, e.g. {aff1,aff2}
]{\fnm{Vivek} \snm{Sharma}}
\author[
   addressref={medialab, mit},                   % id's of addresses, e.g. {aff1,aff2}
]{\fnm{Abhishek} \snm{Singh}}
\author[
   addressref={medialab, mit},                   % id's of addresses, e.g. {aff1,aff2}
]{\fnm{Rachel} \snm{Barbar}}
\author[
   addressref={pcf, medialab, mit},                   % id's of addresses, e.g. {aff1,aff2}
   corref={medialab},                       % id of corresponding address, if any
   email={https://web.media.mit.edu/$\sim$raskar/}   % email address
]{\fnm{Ramesh} \snm{Raskar}}

%%%%%%%%%%%%%%%%%%%%%%%%%%%%%%%%%%%%%%%%%%%%%%
%%                                          %%
%% Enter the authors' addresses here        %%
%%                                          %%
%% Repeat \address commands as much as      %%
%% required.                                %%
%%                                          %%
%%%%%%%%%%%%%%%%%%%%%%%%%%%%%%%%%%%%%%%%%%%%%%

\address[id=pcf]{%                           % unique id
  \orgname{PathCheck Foundation}, % university, etc               %
  \postcode{02139}                                % post or zip code
  \city{Cambridge},                              % city
  \cny{USA}                                    % country
}
\address[id=medialab]{%
  \orgname{MIT Media Lab},
  \postcode{02139}
  \city{Cambridge},
  \cny{USA}
}
\address[id=mit]{%
  \orgname{Massachusetts Institute of Technology},
  \postcode{02139}
  \city{Cambridge},
  \cny{USA}
}
\address[id=harvardmed]{%
  \orgname{Harvard Medical School},
  \postcode{02115}
  \city{Boston},
  \cny{USA}
}
\address[id=harvardash]{%
  \orgname{Ash Center for Democratic Governance and Innovation, Harvard Kennedy School},
  \postcode{02138}
  \city{Cambridge},
  \cny{USA}
}
\address[id=ITGH]{%
  \orgname{Institute for Technology and Global Health},
  \postcode{02139}
  \city{Cambridge},
  \cny{USA}
}
\address[id=NIH]{%
  \orgname{National Human Genome Research Institute, National Institutes of Health},
  \postcode{20892}
  \city{Bethesda},
  \cny{USA}
}
\address[id=UCB]{%
  \orgname{University of California, Berkeley School of Law},
  \postcode{94720}
  \city{Berkeley},
  \cny{USA}
}
% \address[id=UCSD]{%
%   \orgname{University of California, San Diego School of Medicine},
%   \postcode{92092}
%   \city{San Diego},
%   \cny{USA}
% }
\address[id=stonybrook]{%
  \orgname{Renaissance School of Medicine, Stony Brook University},
  \postcode{11794}
  \city{Stony Brook},
  \cny{USA}
}

%%%%%%%%%%%%%%%%%%%%%%%%%%%%%%%%%%%%%%%%%%%%%%
%%                                          %%
%% Enter short notes here                   %%
%%                                          %%
%% Short notes will be after addresses      %%
%% on first page.                           %%
%%                                          %%
%%%%%%%%%%%%%%%%%%%%%%%%%%%%%%%%%%%%%%%%%%%%%%

\begin{artnotes}
%\note{Sample of title note}     % note to the article
\note{\textsuperscript{1}PathCheck Foundation, 02139 Cambridge, USA.\\ \textsuperscript{2}MIT Media Lab, 02139 Cambridge, USA.\\ \textsuperscript{3}Massachusetts Institute of Technology, 02139 Cambridge, USA.\\ \textsuperscript{4}Harvard Medical School, 02115 Boston, USA.\\ \textsuperscript{5}Ash Center for Democratic Governance and Innovation, Harvard Kennedy School, 02138 Cambridge, USA.\\ \textsuperscript{6}Institute for Technology and Global Health, 02139 Cambridge, USA \\ \textsuperscript{7}National Human Genome Research Institute, National Institutes of Health, 20892 Bethesda, USA \\ \textsuperscript{8}University of California, Berkeley School of Law, 94720 Berkeley, USA.\\ \textsuperscript{9}Renaissance School of Medicine, Stony Brook University, 11794 Stony Brook, USA} 
% \note[id=n1]{Equal contributor} % note, connected to author
\end{artnotes}

\end{fmbox} % comment this for two column layout

%%%%%%%%%%%%%%%%%%%%%%%%%%%%%%%%%%%%%%%%%%%%%%
%%                                          %%
%% The Abstract begins here                 %%
%%                                          %%
%% Please refer to the Instructions for     %%
%% authors on http://www.biomedcentral.com  %%
%% and include the section headings         %%
%% accordingly for your article type.       %%
%%                                          %%
%%%%%%%%%%%%%%%%%%%%%%%%%%%%%%%%%%%%%%%%%%%%%%

\begin{abstractbox}

\begin{abstract} % abstract
The COVID-19 Pandemic has left a devastating trail all over the world, in terms of loss of lives, economic decline, travel restrictions, trade deficit, and collapsing economy including real-estate, job loss, loss of health benefits, the decline in quality of access to care and services and overall quality of life. Immunization from the anticipated vaccines will not be the stand-alone guideline that will help surpass the pandemic and return to normalcy. Four pillars of effective public health intervention include diagnostic testing for both asymptomatic and symptomatic individuals, contact tracing, quarantine of individuals with symptoms or who are exposed to COVID-19, and maintaining strict hygiene standards at the individual and community level. Digital technology, currently being used for COVID-19 testing include certain mobile apps, web dashboards, and online self-assessment tools. Herein,  we look into various digital solutions adapted by communities across universities, businesses, and other organizations. We summarize the challenges experienced using these tools in terms of quality of information, privacy, and user-centric issues. Despite numerous digital solutions available and being developed, many vary in terms of information being shared in terms of both quality and quantity, which can be overwhelming to the users. Understanding the testing landscape through a digital lens will give a clear insight into the multiple challenges that we face including data privacy, cost, and miscommunication. It is the destiny of digitalization to navigate testing for COVID-19. Block-chain based systems can be used for privacy preservation and ensuring ownership of the data to remain with the user. Another solution involves having digital health passports with relevant and correct information. In this early draft, we summarize the challenges and propose possible solutions to address the same. 
\end{abstract}

%%%%%%%%%%%%%%%%%%%%%%%%%%%%%%%%%%%%%%%%%%%%%%
%%                                          %%
%% The keywords begin here                  %%
%%                                          %%
%% Put each keyword in separate \kwd{}.     %%
%%                                          %%
%%%%%%%%%%%%%%%%%%%%%%%%%%%%%%%%%%%%%%%%%%%%%%


% MSC classifications codes, if any
%\begin{keyword}[class=AMS]
%\kwd[Primary ]{}
%\kwd{}
%\kwd[; secondary ]{}
%\end{keyword}

\end{abstractbox}
%
%\end{fmbox}% uncomment this for twcolumn layout

\end{frontmatter}

%%%%%%%%%%%%%%%%%%%%%%%%%%%%%%%%%%%%%%%%%%%%%%
%%                                          %%
%% The Main Body begins here                %%
%%                                          %%
%% Please refer to the instructions for     %%
%% authors on:                              %%
%% http://www.biomedcentral.com/info/authors%%
%% and include the section headings         %%
%% accordingly for your article type.       %%
%%                                          %%
%% See the Results and Discussion section   %%
%% for details on how to create sub-sections%%
%%                                          %%
%% use \cite{...} to cite references        %%
%%  \cite{koon} and                         %%
%%  \cite{oreg,khar,zvai,xjon,schn,pond}    %%
%%  \nocite{smith,marg,hunn,advi,koha,mouse}%%
%%                                          %%
%%%%%%%%%%%%%%%%%%%%%%%%%%%%%%%%%%%%%%%%%%%%%%

%%%%%%%%%%%%%%%%%%%%%%%%% start of article main body
% <put your article body there>

%%%%%%%%%%%%%%%%
%% Background %%
%%

\section{Introduction and Related Work}
COVID-19, a global pandemic has high transmission and mortality rate, especially among the elderly, healthcare workers and those with underlying comorbidities. With the vaccines still undergoing final approvals, the mainstay of pandemic management depends on key public health intervention including early detection,  mitigation through isolation and contact tracing. Countries across the world are taking strict measures to manage the overwhelming and unprecedented impact of the pandemic Digital health technology, the mainstay of pandemic strategy can achieve superior outcomes that are difficult to achieve otherwise. Integrating these digital tools into public health interventions has revolutionized surveillance, testing, contact tracing, healthcare access  and quarantine measures aimed at control and mitigating COVID-19 spread thereby aiming to reduce COVID-19 infection and deaths
Big data and artificial intelligence (AI), used widely in healthcare, have also helped in tracking people and thereby reducing the spread of infection. Migration maps including mobile phones, mobile payment applications and social media are used to collect real-time data on the location of people. Machine learning and disease spread models were developed to predict the regional transmission of the virus that help guide policy and containment measures. Efficient use of big data has contributed to reduced spread and early detection, thereby saving thousands of lives. The need to track the spread of COVID-19 has fuelled the innovation of data containing dashboards that display variables such as disease burden including number of cases, deaths, recovered, and active patients.s. Some dashboards could plot trends according to demographic and geographical characteristics. The limitations of AI need to be addressed and requires training with COVID-19 datasets. The accuracy, validity and reliability of these digital tools are a concern.
This viewpoint provides an outline for the application of digital technology in the landscape of pandemic management and response. Also, highlighting the challenges and success of various digital technologies used by countries to manage the pandemic. 

Since December 2019, the COVID-19 pandemic has contributed to an exponential growth in the development of diagnostic technologies and interdisciplinary and inter-country scientific data sharing. A recent bibliometric study has shown that the availability of evidence in response to the COVID-19 pandemic has been much more efficient than other recent epidemic events, such as the 2015–16 Zika virus epidemic and the 2014–16 Ebola virus outbreak due to technology and science advancements and rapid sharing of information. More than 2,500 articles related to COVID-19 were published in the first 4 months of the pandemic, compared to only 88 articles related to both Zika and Ebola viruses in the same epidemiological period. \cite{1}

The pace at which viral genomic sequences have been available to the public during the COVID-19 pandemic also illustrates the rapid pace of data sharing over time. \cite{2} As of 31st December 2019, 19 gene sequences of the SARS-CoV-2 virus were already obtainable through the GISAID database (gisaid.org), which now has over 40,000 viral genome sequences shared by laboratories around the globe. As a comparison, it took almost 3 years for the number of sequential viral genomes to reach 1,500 sequences during the Ebola virus outbreak. The rapid developments of SARS-CoV-2 genomic sequences contributed to both the rapid development of the gold standard molecular diagnostic tests for COVID-19 and to the development of simplified protocols for complete viral genome sequencing and analysis and lab-based serology assays using recombinantly produced SARS-CoV-2 proteins.

In this early draft, we aim to understand the digital landscape of the current testing scenarios. We go to the granularity of understanding the User agency and control, understanding the digital workflow of testing, how various organisations (or governments) are using the digital solutions for testing.  Furthermore we look into understanding the major challenges in the current workflow including but not limited to - privacy, security, miscommunication, trust and laws. We look at various case studies across countries to understand the effects of different digital solutions and leave an open thread for the solutions.

% \subsection{User Agency and Control of Data}
% One of the key challenges in the digital management of testing has been how the agencies acquire and control the data or the users for their use and the privacy policies they follow. Confidentiality has been one of the most important issues and concerns of the digital revolution and given rapid use of digital platforms on COVID-19 management. Increased privacy concerns are attributed to several health platforms requiring patients to share their personal protected health information, including the patient’s name, phone number, email, race, country of birth, SSN, residential address and even travel history. This is problematic because of the lack of policies and regulations about the use of personal data [Link]. Further, due to the usage of sensitive data, users face the security risk of being tracked, man-in-middle attack and data snooping by a third party. 

% Thus, it becomes very important to have the right policies set which inform the user about the need for their data and consent of use from the patient is a must. 

\subsection{User Agency and Control of Data}
One of the key challenges in the digital management of testing has been how the agencies acquire and control the data or the users for their use and the privacy policies they follow.  Confidentiality has been one of the most important issues and concerns of the digital revolution and given rapid use of digital platforms on COVID-19 management. Increased privacy concerns are attributed to several health platforms requiring patients to share their personal protected health information, including but not limited to the patient’s name, phone number, email, race, country of birth, SSN, residential address and even travel history. This is problematic because of the lack of policies and regulations about the use of personal data. \cite{3} Further, due to the sensitive nature of the data, users face the security risk of being tracked, man-in-middle attack and data snooping by a third party. 

Thus, it becomes very important to have the right policies set which inform the user about the need for their data and consent of use from the patient is a must. 

\subsection{How other epidemic tests are managed (Ebola, HIV, malaria, Swine flu, zika, etc)}
The Zika outbreak in 2016 was a global public health concern and digital technology played a vital role alongside traditional public health approaches. Mathematical modelling helped in understanding the Aedes spp. Vectors in terms of transmission, host-virus interaction, and effectiveness of  potential interventions. Geographical information systems (GIS) including epidemiological maps were used to estimate magnitude, transmission and spread of the Zika outbreak. The category of big data included social media analytics and web-based surveillance particularly on the OpenZika project, Digital Participatory Surveillance (DPS), and ProMED. DPS is a form of digital surveillance where the public reports signs and symptoms through a structured electronic form and the data is summarized via epidemiological maps. Zika tracker, a smartphone-based diagnostic platform could be used anywhere in the world to report suspected or confirmed cases of Zika. WHO`s Zika App is a similar app that provides real-time data to all users. Researchers in Brazil developed MinION that is a novel real-time genome-sequencing device. An insecticide wearable device that releases Metofluthrin (traditional repellant of Aedes aegypti) reduced the number of mosquitoes attracted. \cite{4}

During the Ebola outbreak, diagnosis was difficult especially in asymptomatic people. The gold-standard diagnostic test for Ebola virus, the real-time RT-PCR needed substantial infrastructure, operation and maintenance of complex equipment, and expertise in molecular techniques. Limitations of the diagnostic tests were addressed by digital solutions. The Democratic Republic of Congo (DRC) used the National Health Information System to create a dashboard called PATH for monitoring Ebola cases. The US Center for Disease Control and Prevention (CDC), World Health Organization (WHO) and other organizations worked together to develop digital maps and graphs. In 2019, the Ministry of Health of DRC launched a national agency for clinical information and health informatics (ANICiiS). During the 2014-2016 epidemic, mHERO, an SMS system was set-up to share information with health workers. UNICEF also launched the Rapidpro free source platform that provided SMS messaging in Sierra Leone where it was rare to possess a smart-phone, internet, or even phone credit. \cite{5}\cite{6}\cite{7}

HIV/AIDS control strategies included digital solutions such as  mapping and maintaining surveillance of high risk behaviors, individual level intervention that targets key populations (during a low-level epidemic), providing voluntary testing and counseling (VTC) and screening for sexually transmitted infections (STIs).

Malaria remains endemic in several countries despite multiple interventions. Containing malaria outbreaks remained a challenge because it was difficult to predict who would contract the disease. As a result, resources were not deployed in the correct region. This caused the outbreak to grow. A quick and reliable digital solution was required. The National Aeronautics and Space Administration (NASA) used satellite data to help forecast malaria outbreaks by identifying areas of prime breeding grounds for these mosquitoes that overlapped with human population. They use the information on precipitation, temperature, soil moisture, and vegetation around the world and feed it into a prediction model and tell where a puddle is most likely to form. Simply mapping where people reported malaria is not enough as the place of infection is usually different. They also track humans to see where most people are getting infected i.e hotspots. But for the purposes of predicting a malaria outbreak, the map did not tell a complete story. They could successfully make projections down to the household level, allowing for resources to go where they’re needed. The population models could be used for tracking not only malaria, but Zika and Dengue as well. \cite{8} The Digital Solutions for Malaria Elimination (DSME) Community of Practice (COP) helped in building malaria surveillance systems that made accurate data reporting easier and thus improved decision-making processes. \cite{9}

\section{Testing Workflows}
\subsection{User experience}
The rapid spread of COVID-19 has been worrisome for many individuals across the globe with respect to the safety of security, workflows and measures being taken to overcome these issues. There is an increasing emphasis on providing great user experience and information to the citizens so they stay informed and aware of the recent developments, policies, precautions and measures that have been taken up to fight the virus.
\subsubsection{User engagement summary table}
In this table, we highlight the similarities and differences in user engagement within public/government sites, private clinics, employee/campus organizations, and entering venues (such as flights). It is important to understand the different facilities which are provided by the different organizations and institutions and how the public and government sites can be modified more to support the facility of giving out reminders and covering the insurances for the patients. Also the entry pass system at the public/ government sites should be more digitised to avoid the issues of data privacy and spoofing. Adding on, it is most important to provide the scheduling facility across all the different organizations to avoid the issue of overcrowding and have a systematic approach in place. Even the test results should be supported with a detailed report covering points ranging from a detailed analysis of virus to the precautionary measures that need to be taken by the patient to avoid getting infected in future, in case they are affected a report needs to be shared to indicate the exact steps the patient should take in order to stay protected and at the same time not transmit the infection to others.

\begin{figure}[!h]
    \includegraphics[width = .95\linewidth]{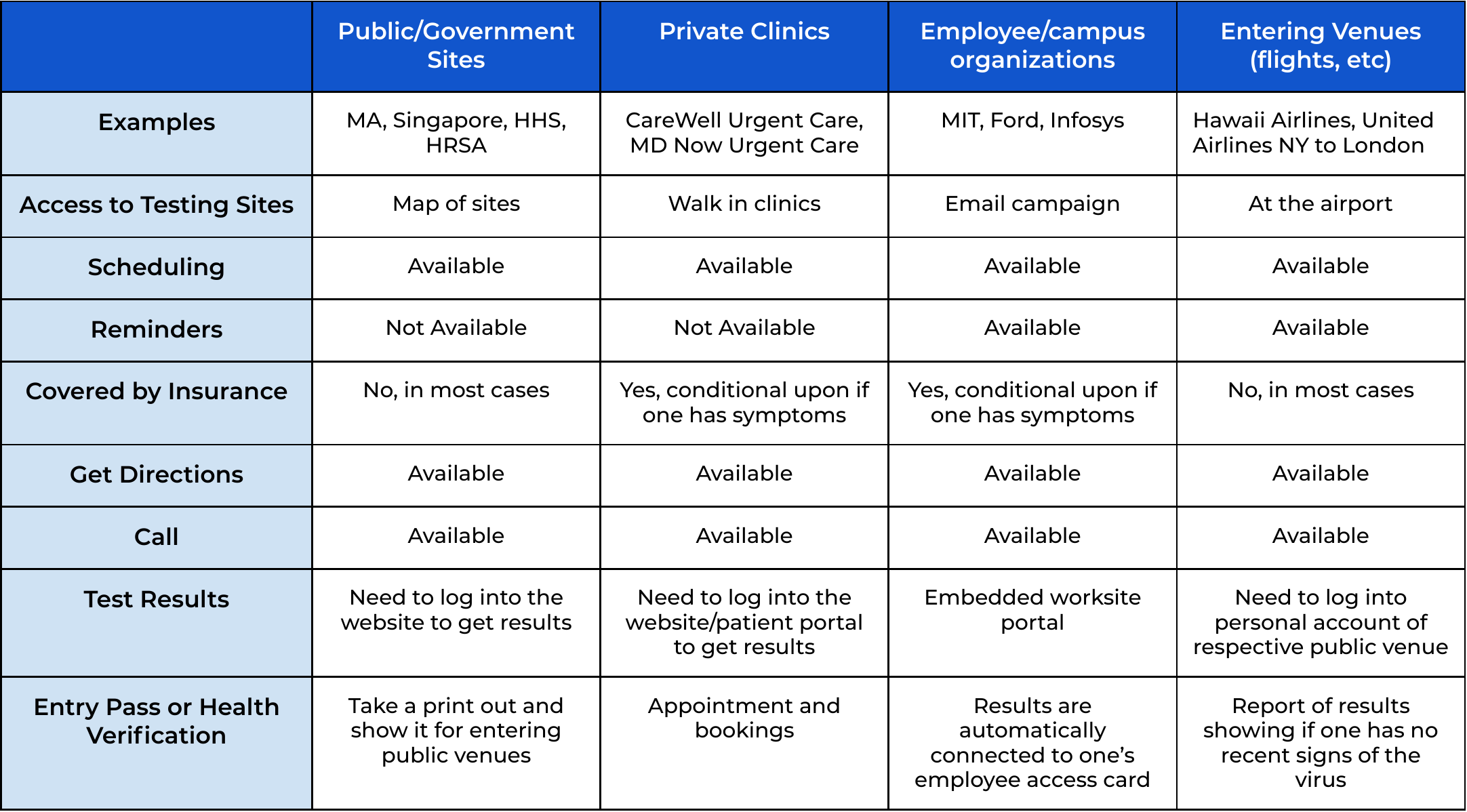}
    \centering
    \caption{User engagement summary table}
    \label{fig:user_engagement}
\end{figure}

\subsubsection{User Journey Table}
It is also important to understand more about the user’s journey and the steps taken by them in order to get tested. First via social media campaigns, televisions and youtube videos the end user is educated about the different COVID-19 testing centres and sites. Using this information the user tries to explore more about the different sites which may be public, private, educational sites and draws a comparative analysis between them. Once they are certain about the type of site/ organisation they would want to visit next they analyse the different centres and hospitals available for them to visit and this decision is taken based on several parameters such as the type of testing available, cost of testing, time to get the results and one of the crucial points would be the location of thee testing sites. Lastly, once tested the patient/user tends to receive more updates and reminders on taking the necessary dosage and also receive timely reminders via messages.\\ 

\begin{figure}[!h]
    \includegraphics[width = .95\linewidth]{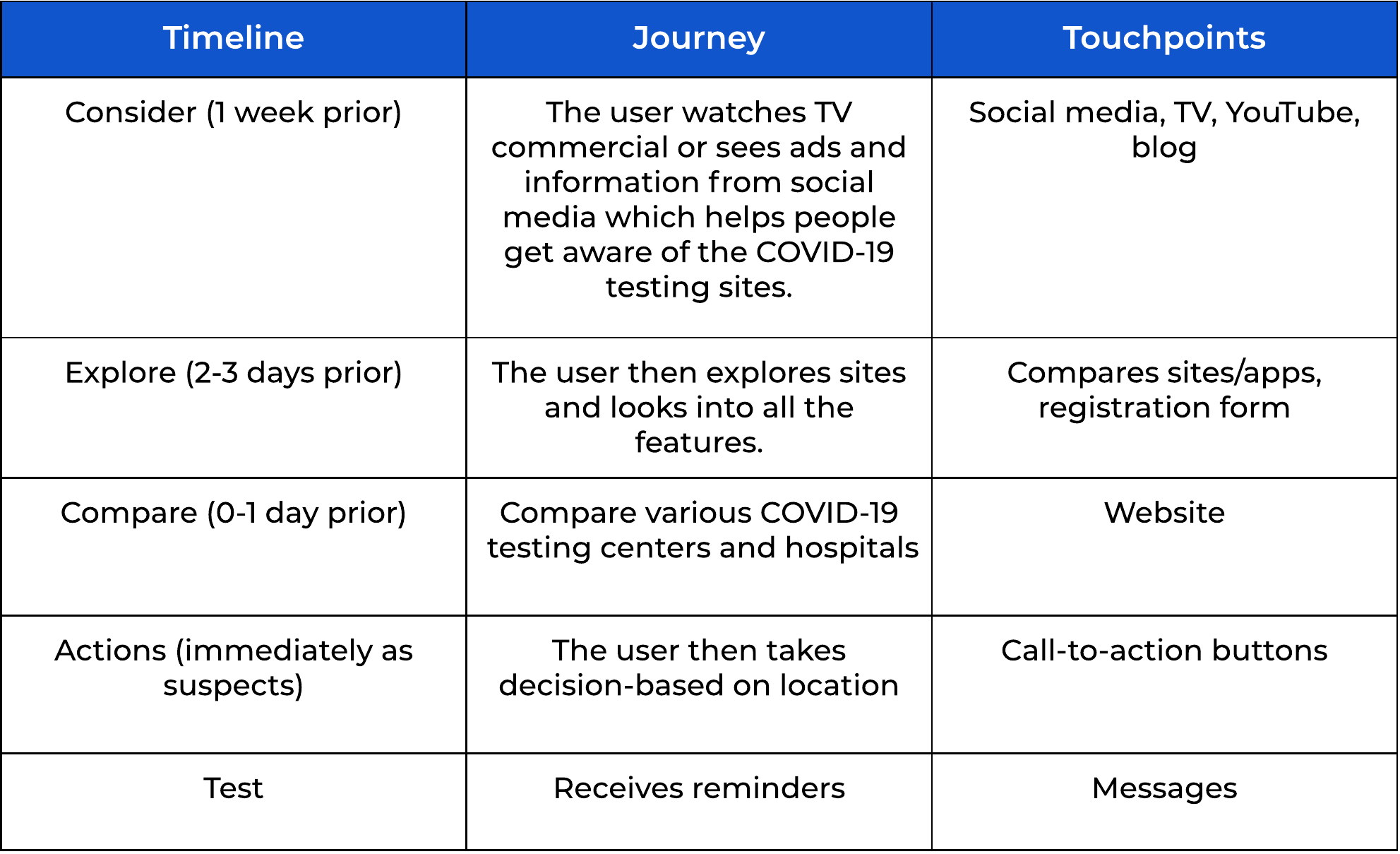}
    \centering
    \caption{User journey table}
    \label{fig:user_journey}
\end{figure}

\subsubsection{Test site logistics (LIMS systems, Lab Info Management Systems) (Quest, LabCorp, Verily, etc)}

\begin{figure}[!h]
    \includegraphics[width = .95\linewidth]{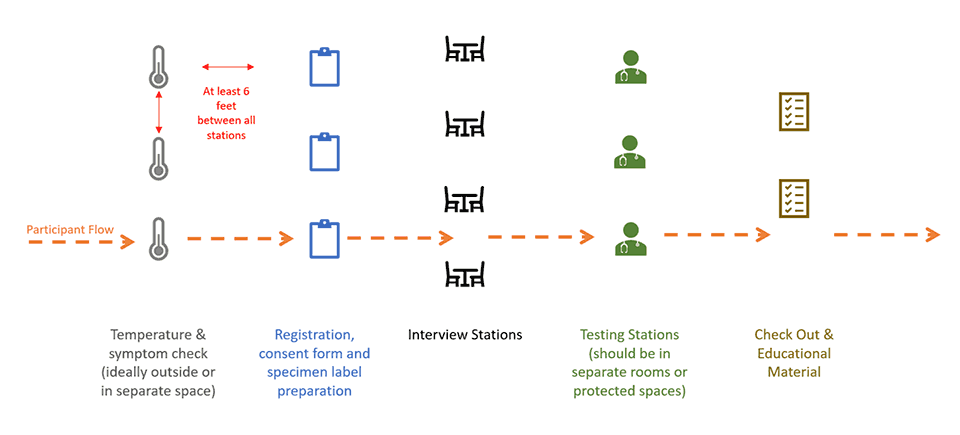}
    \centering
    \caption{CDC recommended COVID-19 Flow Screening Process \cite{cdcflowscreening}}
    \label{fig:flow_screening}
\end{figure}
Figure \ref{fig:flow_screening} illustrates the standardised process flow for COVID-19 testing. The participants should flow in one direction and each of the stations should be separated by at least 6 feet. Participants begin with a temperature and symptom check. This is also where participants are given a face covering if they don’t already have one. Then, they fill out all the necessary paperwork (pens and clipboards are disinfected in between participants). Participants proceed to the interview stations where they are screened for symptoms, after which they are put in testing stations where they are given a swab to collect their specimen. The specimen is then put into a vial and dropped into a transport bag. Lastly, participants are then given the information needed to receive their results \cite{10}.

There are many software and data management systems in use as well. Data management systems are essential during COVID-19 case investigations and contact tracing. This data is stored in different ways based on the guidelines of the local jurisdiction or testing sites, but typically is stored in storage platforms like cloud. It is important for the data to be stored in a way that preserves anonymity and protection of Personal Health Information (PHI). In addition, Lab Information Management Systems (LIMS), a type of software present in most modern laboratories, can be used at COVID-19 testing sites. LIMS could improve productivity and efficiency of laboratories by increasing the speed of test processing, reducing costs by automating, and streamlining lab workflows.

Test sites should be in large open areas (preferably outside) to allow for effective social distancing. It is also important for test sites to have centers to isolate patients who test positive for COVID-19. In a test site, keeping staff members safe is extremely important. The CDC recommends gloves and face masks for all staff members present at a test site, and gowns and N95 (or equivalent) masks for anyone who will be within 6 feet of any person being tested. The person being tested should also wear a face mask. Aggressive sanitation is necessary, including changing gowns, respirators, and face masks if they are soiled. In addition, these should be replaced in case they are accidentally touched or if they come in close contact with a person being tested.

\subsubsection{Dissemination of results to the user (website, app, phone call, email, text) (Comcare, Salesforce)}

Testing forms a crucial point of intervention for mitigating the COVID-19 pandemic. The results of one’s COVID-19 test along with other relevant information will assist one’s healthcare provider to make informed decisions regarding patient’s health including recommending treatment and mitigation strategies. Timely and proper communication of results will also help in limiting the spread of COVID-19 to one's family and the larger community. The mechanism of disseminating results varies greatly across countries and can be through web based secure file transmission, an SMS, phone call, email or within an embedded app portal. Digital or electronic-based reporting would reduce unnecessary travel and risk of exposure. The timeframe for tests to be made available depends upon the type of test and the area it was done. The process for Rapid antigen tests is faster than RT-PCR. Point of care testing in community settings like schools, college campuses, nursing homes and employer organisations remain crucial for public health mitigation strategies.Testing sites ,generally employ the following methods for results notification:
\begin{itemize}
    \item Electronic laboratory reporting:It refers to automated messaging of reports sent using one or more electronic communication portals like SMS, email etc. ELR ensures that health departments and patients receive timely, accurate, complete and consistent information and improve the deployment of public health responses to outbreaks and cases of notifiable conditions. 
    \item Point of Care-Test reporting: it refers to rapid testing where results can be provided within minutes of the test being administered, thus, allowing for rapid decisions about patient care. However, results of these tests cannot be completely said to be reliable.The test results are also required to be notified to local and state health departments.
    \item Through fax the test results should include:test performing facility name, type of test used, explanation of results,patients particulars etc.
\end{itemize}

Here, below are few of the COVID-19 testing sites by public , private and employer organisations  to illustrate test results dissemination process: 

\begin{itemize}
    \item \textbf{NHS COVID-19 (Public or gov. testing site)}: It is a U.K. government testing site \cite{11}, funded by the  Department of Health and Social Care (DHSC). Test results are communicated through text or email. If one uses the NHS COVID-19 -19 app, tests are accessible there only. The results are provided the very next day but at most, it may take up to 3 days. If one tests  positive, one may be contacted by the NHS and asked for further information
    \item \textbf{MIT COVID-19 Pass (Campus Testing Site)}: Covid Pass is an application for the MIT community members to use in order to complete requirements for entering the MIT campus and using the resources which they have access to \cite{12}. In order to gain access to the campus, one must agree to conditions like daily health attestation, and regular COVID-19 testing. Testing frequency is contingent on how often the person is on campus. COVID-19 Pass will notify ( through the email and mobile push notification) about testing status early each morning. This will help to keep track of when individuals need testing. Test results are generally provided within 1-2 days which will be available in COVID-19 Pass App on MIT Atlas and not in HealthElife. Though, tests done through appointments will have their results visible in HealthElife. If an individual does not yet have full authorization to use the COVID-19 Pass app, they may still see results at covidpass.mit.edu using MIT Kerberos authentication. A physical document of the result can also be requested by call at MIT Medical COVID-19 hotline. If tested positive, MIT Medical will contact the individual directly and the contact tracing process will begin. If no communication is made within 48 hours, we can presume tests are negative. Amidst a global pandemic, the MIT COVID-19 Pass is an efficient and time saving digital solution. In the hassle and rush of getting work done in the morning, when someone may forget that they need to be tested, the COVID-19 Pass app will notify them.
    \item \textbf{Minnesota Department of Health}: It is a public health department that provides community testing \cite{13}. If the test comes negative, one will receive a text and/or email notifying that one can access results online. If there is no access to the internet, there will be a number on which calls can be made for results. In case it is positive, one will receive a phone call from the triage line. 
    \item \textbf{Henry Ford Health System}: The Henry Ford Health System has provided an online COVID-19 screening tool to assess risk \cite{14}. Remote monitoring and virtual consultation in the form of E-visits and video visits has been made possible. The Henry Ford Health System has provided an online COVID-19 screening tool to assess risk. Remote monitoring and virtual consultation in the form of E-visits and video visits has been made available. 
    \item \textbf{Carewell Urgent Care}: Carewell \cite{15} is a community-based health care system that provides both patients and employers COVID-19 testing services.For employers, results will be communicated through fax. For others,  in the case of the PCR test, results are displayed on the Quest Diagnostics Website while antigen test results are posted on carewell urgent care website itself. 
The dissemination of results is a crucial step in the user experience since it plays an important role to make sure the user is aware of the precautionary measures to be taken, will help make more informed decisions and also keep them updated about the current testing workflows and practices.

\end{itemize}

\subsubsection{Efficacy, ethical and privacy concerns}

An immunity passport based on vaccine-induced immunity would eliminate some of the moral hazards and ethical concerns raised by experts: that some non-infected individuals would see it as an incentive to expose themselves to the virus in order to develop antibodies [ref] (needs fix) as a way of receiving immunity passports and practice herd immunity. There’s also a concern people would be penalized for low-risky behavior resulting in a lack of exposure to the virus, which would make them ineligible for a card based on infection-related immunity.
 
Yet, other efficacy, ethical and privacy concerns remain relevant to digital passports based on vaccine-induced immunity. Evidence suggests \cite{19} that, despite being immune, an individual might still transmit disease and infect others. Though different places are planning for non-discriminatory and equal access vaccine distribution, there are multiple unknown challenges along these lines. This could hinder the effectiveness of digital immunity passports . Otherwise, according to ethicists, digital immunity passports might increase racial and social inequalities \cite{22} already exacerbated by the pandemic, repeating the history of immunity passports during  “yellow fever”  epidemics at the beginning of the 20th century. These were used as a means of racial discrimination, and oppression. Digital immunity passports could also facilitate social exclusion \cite{23} and discriminatory work environments, where “an official stamp of immunity to COVID-19, or personal willingness to risk the disease” would “... become a prerequisite for employment”. \cite{22}
 
Finally, the immunity certificates’ credentials could create privacy and security risks. And the centralized nature of a digital credential system, where a single institution would be responsible for data collection and credential management, could lead to privacy violations and abuse of power. Dakota Greuner from Harvard Edmond J. Safra Centre for Ethics’ COVID-19 Rapid Response Impact Initiative argues \cite{18} that for any digital health passports to be effective, it should possess three features:
\begin{enumerate}
    \item Be \textbf{privacy-preserving} - Personal data should only be managed by the user, stored on the device, and shared only after explicit user consent. It also should be encrypted and secured by biometrics. Selective disclosure of identifiers should be incorporated in the design (for example, age but not birth date etc.) \cite{18}
    \item Be \textbf{portable and widely adopted} – It should be based on open standards, so it is broadly usable across systems and devices.
    \item Be \textbf{trustworthy} – This could be achieved with a digital signature of authenticity secured by the use of cryptography, so that information on the date and the credential could not be changed or hacked.
\end{enumerate}

\textbf{Decentralized systems of credentials deployed at a large scale} are being suggested as a  “privacy-protecting model of immunity certificates.” \cite{18} Owing to the decentralised and trustworthy framework, block chains have come across as a solution to deliver critical information in a secure and privacy preserving way. 

With the nature of health records and issues like record linkage, privacy and misinterpreted results (due to false positives or miscommunication) it becomes difficult do extract insights from them. This coupled with the lack of trust (as health information is very personal) makes it difficult to have data shared. This results in data silos and many missed opportunities of information sharing and understanding of health information. Blockchain based solutions can help to mitigate some of these issues as the data ownership is still will the respective person (or hospital) and it ensures security and reliability.
A few of the impactful applications for COVID-19 response which are contact tracing, sharing and handling patients data employ the use of Blockchain.

\subsubsection{Digital Health Verification}
From Ireland to Chile, efforts are underway worldwide to introduce health immunization passports as a way of easing out social distancing rules and allowing for individuals to go back to work, travel or go out to public venues, such as restaurants and sports events \cite{16}\cite{17}. These digital immunization passports are meant to be a digital equivalent of paper certificates \cite{18}, indicating the holder is at low risk of transmitting the coronavirus, providing  evidence of acquired immunity to the virus, by having recently tested negative to the virus, having antibodies or been vaccinated.  Immunity passports have been compared to international certificates of vaccination, such as the “Carte Jaune” for yellow fever. However, there are significant differences between the two types of documents to be noted. Such credentials could take the form of a wristband / wearable device, a digital certificate via smartphone application \cite{19} or a biometric ID card, among others. The majority of initiatives focus on digital credentials that could be carried on users’ smartphones (for instance in the form of the QR code).  Individuals with an immunity passport could be exempt from physical restrictions and could return to work, school, and daily life. However, immunity passports pose considerable risks including, by not limited to, scientific risk, practical risk, equitable risk and legal challenges. Considering all these risks , the users should be wearing masks and follow minimum safety standards. 

The immunity passports rely on two concepts: \textbf{infection-related immunity} and \textbf{vaccine-induced immunity}. Both raise issues of “the degree of immunity induced and the duration of immunity” \cite{19}. WHO is not recommending \cite{20} an introduction of infection-related immunity passports based on “the lack of sufficient evidence that people with COVID-19 antibodies are immune to new COVID-19 infections.” \cite{21} Vaccine-induced immunity as a basis for immunity passports is more promising, with recent results in animal trials revealing a correlation between vaccines and protection. A uniform vaccination, experts say, could provide a more “predictable pattern and duration of immunity.”

\subsubsection{Examples of digital health passports}
As early as April 2020, countries and companies have started to look into digital health passports based on potential infection-related immunity. They did so in an effort to speed up economic recovery and help people to come back to work, travel or participate in sports. These and others are expected to be expanded to certify vaccination once it is available. 

The list of available solutions is growing day-by-day with companies rushing to conquer the safe-back-to-work market, including IBM Watson Digital Health Passport \cite{24}, VHealth Passport by VST Enterprises \cite{25}, WIShelter by WISeKey \cite{26}, COVI-Pass by Tento Health \cite{covipass}, and other solutions introduced by German IDNow and UK companies Onfido and Yoti.  

\begin{enumerate}
    \item \textbf{Chile released certificates}: Chile has started a roll out of smartphone-based “release certificates” instead of “immunity passports,” currently valid for 3 months and shared in the form of the QR codes \cite{27} issued to people that did not have any symptoms for 14 days after testing positive for COVID-19. 
    \item \textbf{Estonia digital immunity passport}: In Estonia \cite{28}, a digital immunity passport developed by the “MTÜ Back to Work” non-profit  formed by tech companies such as Transferwise, Guardtime or and Bolt. Similarly, to a Chilean solution, users share the information about their test results and medical certificates via a QR-code generated after digital authentication. 
    \item \textbf{Health Passport Ireland}: Ireland began a digital health passport pilot in August \cite{29}. It is based on privacy-preserving technologies, with only an authorized medical administrator allowed to create Health Passport Ireland accounts and make updates of the test results. The app displays green status if the test result was negative, red if positive and amber if a new test is required \cite{16}. The color-code certification system is somewhat similar to the apps that have been introduced in China \cite{17} and the UAE’s Alshosn app \cite{30}. 
    \item \textbf{COVID-19 Vaccination Certificate pilot}: Most recently, WHO established an international partnership between Estonia on WHO Digital COVID-19 Vaccination Infrastructure \cite{31} based on blockchain technology for a decentralized and secure data management and authentication of an digital International COVID-19 vaccination certificate based on the vaccine-related immunity. The system, based on a web platform and an app, will be tested via a 12-week pilot in Estonia, and if successful, it might become a standard of the future when the vaccine is available on a large-scale and vaccine-immunity is proven to be effective.
    \item \textbf{COVID Credentials Initiative}: The COVID-19 Credentials Initiative (CCI) \cite{32} is an international community of more than 300 members from over 100 organizations hoping to deploy and/or help deploy privacy-preserving verification-based credential projects in order to reduce the spread of COVID-19 and strengthen our societies and economies. The CCI has launched the usage of digital identity in order to curb the spread of COVID-19. Their aim involves developing “immunity passports” and much more. In this, The CCI uses verifiable credentials. Inspired by the functional utility of a physical credential (e.g. the cards in one’s wallet), a verifiable credential (VC) is a declared assertion that contains documented claims about a person or a particular organization. The niche of VCs that make them so pivotal is that they're digitally native and cryptographically secure, providing a great privacy-preserving alternative to other types of credentials, if used responsibly. Upon accepting a VC (e.g. a driver’s license) from a trusted issuer (e.g. a government body), holders can generate proof with minimal information (e.g. over 18) to validate to a verifier (e.g. a liquor store  ) that they posses their own VCs with specific information (e.g. age), qualifying them for particular types of access defined by the verifier. And there is no need for direct contact between the issuer and verifier throughout the process. 
    \item \textbf{Travel health passports}: Airlines and tourism organizations are also introducing digital health passports for travelers to enable safer cross-border travel. United and Cathay Pacific \cite{United_and_cathay} are piloting a program called CommonPass \cite{common_pass}, a privacy-by-design, globally interoperable platform developed by The Commons Project in collaboration with The World Economic Forum and other public and private partners. It allows individuals to access via app their test results. In the future, they also will be able to access their vaccination records, available from trusted sources \cite{common_pass_site} and give consent to use those records to validate their COVID-19 status. ClearMe, a biometric identification solution developed by Secure Identity LLC, currently used by Delta and United Airlines as well as numerous sporting facilities, includes Health Pass \cite{health_pass} as a free feature. It allows for real-time screening for possible symptoms via surveys on the app and temperature-checks at kiosks installed at the venues and generates a QR-code with a verified identity and health-check status. In July, the Tourism Data Driven Solutions (TDDS), a Spanish software company, developed Hi+ Card \cite{hi_card}, a mobile app supported by the UN World Tourism Organization (UNWTO), that is meant to serve as an international health identification card. The solution uses blockchain technology and encryption and is General Data Protection Regulation (GDPR) compliant.
    
    Digital Health Verification is one of the most factors that needs to be addressed to ensure the ethical, political and social concerns.  The health passports might not eliminate the threat completely, but is an essential tool in this pandemic. Furthermore, these health passports should be privacy preserving, trustworthy and widely accepted. It is also important to be mindful of the potential social and racial discrimination that might entail a health passport based solution.
\end{enumerate}

\subsection{Dashboards and Tools for Public Health}
Dashboards and other web tools play a very important role in sharing the message about the magnitude, severity of the COVID-19 crisis and its impact on health. There are numerous key testing players which are leading the way and helping significantly with testing and analysing the pandemic situation. Combined with the testing efforts, having a visual representation of the number of country-wise cases along with a wide array of filters to study socio-demographic and geographical variations can be of great use. Dashboards enable us to serve this very purpose. With the purpose of getting a grasp on the COVID-19 pandemic, numerous dashboards have been created at university level, regional, national and global level.

\subsubsection{Dashboard tools for labs}
The COVID-19 diagnostic testing is carried out to check if an individual is infected with the SARS-CoV-2 (the virus that causes COVID-19), a detailed overview about the clinical landscape of testing has been covered by Gandhi et. al \cite{33}. Mainly two tests were aproved by The US Food and Drug Administration (FDA) for detecting of COVID-19, which is the PCR test and the Rapid Antigen test. All the testing laboratories are CLIA certified. CLIA, The Clinical Laboratory Improvement Amendments are United states federal regulations and regulatory standards that are applicable to every clinical laboratory testing, except clinical trials and basic research which are carried out on humans in the United States of America.\cite{34} Testing is of utmost importance in current times and some of the leading players when it comes to testing include Quest Diagnostics \cite{35} which handles approximately 20\% of the testing efforts across the USA with over 40 years of proven effectiveness and experience in handling infectious disease testing, Medpace Central Labs \cite{36} who have been committed to assisting in fighting against the COVID-19 pandemic, Icon \cite{37} who offer a wide array of services right from testing to maintaining a coronavirus observatory which is a powerful dashboard which provides updates regarding the vaccine trails, regulatory updates, diagnostics and much more related to COVID-19 via Twitter, Syneos Health, \cite{38} the only fully integrated biopharmaceutical solutions organizations have been taking extensive measures in helping battle the pandemic. Some other leading names when it comes to testing are Verily, BioReference Laboratories, and Mayoclinic to name a few. Verily \cite{39} has introduced Project Baseline which is designed to support people right from screening through testing at respective community-based testing sites and receipt of individual test results. As of September 2020, 6L people have been tested through the Baseline COVID-19 program. BioReference Laboratories \cite{41} have played a key role in the crisis as well as provided support to various patients as well as government agencies. Mayoclinic \cite{42} which is an American non-profit organization focused on patient care, research and education has also played a significant role in terms of testing. It is crucial to understand about the various individuals who are involved in the testing workflow and how their contribution is vital to the success of preventing the spread of the virus and carrying out tests effectively and optimally.

\subsubsection{Dashboard tools and software for public health, govt, and CDC/WHO}
Dashboards are the landing pages for interactive maps and visuals showing where the virus has spread. It gives  a breakdown of which nations, states, cities, and neighboring areas have positive cases and could predict which regions are likely to see new outbreaks. This helps in predicting future trends and taking action such as putting area restrictions or a complete lockdown. Data dashboards provide latest figures on the number of positive cases (mild, moderate and severe), recoveries,  deaths,  infection rate, and cases per million of the population in that area. While some of them also give resource-related information that includes availability of beds, testing kits, PPEs, masks, and ventilators, dashboards also give diagnostic test-related information including number of tests and  positivity rate.

\textbf{Government}: The World Health Organization (WHO) dashboard is one of the most broadly-accessed and predicated on visualization techniques of the COVID-19 pandemic. The general populace can navigate this dashboard easily by discovering numbers across the world, such as infected, dead, and recoveries. In addition, the WHO dashboard assists users in evaluating COVID-19 data in real-time as graphics change based on the ever-changing situation. It  gives information regarding COVID-19 cases globally in the form of Choropleth and bubble maps. It gives information on new cases, confirmed cases, and number of deaths. The data table covers different regions, states, and areas which also deals with transmission classification. The group demonstrated of COVID-19 on the WHO dashboard provides data on the follow-up guidance for comprehending the pandemic. For example, users can understand the method used to determine the COVID-19 information and to update the dashboard in real-time. The Explorer section directs the public on trying to navigate through the dashboard and offering to help them to make sense of the COVID-19 statistics. 

The Centers For Disease Control And Prevention (CDC) COVID-19 data tracker gives information about the total number of positive cases, deaths, trends including demographics and population factor. It also compares the trend in different states and helps in forecasting future trends for the United States. Interestingly, it also includes community impact of the pandemic in terms of  mobility of individuals and social impact such as news reports on school closures, localized outbreaks, state of emergency declarations, etc. The CDC dashboard persists to be one of the most frequented systems, as the COVID-19 pandemic continues with the community using dashboard updated information to comprehend the virus. Unlike others, the CDC supports a framework of all factors necessary for the Pandemic Assessment, including the dates of infection and the racial groups affected. This dashboard comprises a segment at the top in which users could see a preview of existing COVID-19 data prior to actually moving ahead to other areas. 

\textbf{University level}: The only way to grasp and understand the magnitude of the COVID-19 pandemic for the masses is through visualizing the data pertaining to the crisis. People, in general, can be overwhelmed by huge numbers and may find the patterns contradicting. Visualization of the COVID-19 scenario makes it easier for everyone to analyze and understand the situation in a better way. Experts at all levels, right from university-level to government organizations have published COVID-19 dashboards at global, country, state, and regional levels showcasing an array of parameters right from a comparative analysis of the COVID-19 cases in the previous day or month to the number of recovered patients and much more. At the University level, the dashboard visualized by Amherst College \cite{44} has made much noise by being the only college to receive an A++ rating by We Rate COVID-19 Dashboards \cite{dashboards} which is a panel of academic and medical experts who rate dashboards created by universities and colleges across the country. The top 5 universities who have excelled at providing data of the vicinity via dashboards according to the ratings are Amherst College \cite{44}, Wagner College \cite{45}, Tulane University \cite{46}, Ohio State University \cite{47}, and George Mason University \cite{48} as of 12th November 2020. These dashboards provide an overview of the key metrics for their vicinity. The daily cases, hospital capacity among them, average positive cases, and many other parameters are included, The student/faculty and in the campus cases are also displayed at a daily, weekly, and monthly level along with the number of isolation cases. These parameters combined with the many other parameters can help anyone visiting the vicinity to analyze the scenario and make an informed decision accordingly.

\textbf{National level}: The government, administration and healthcare experts need to keep an eye on the overall COVID-19 situation in the country. A dashboard, at the national level remains the cornerstone to monitor long term impacts and trends, quantify the burden and assist in decision and policy making. The CDC COVID-19 Dashboard \cite{49} is easily one of the most visited dashboards across the United States of America with the public using it to understand the virus better as the pandemic continues to accelerate. The CDC Dashboard stands out amongst other dashboards since it considers all the factors which are necessary for pandemic evaluation like racial groups affected and infection dates as well. The dashboard by Carnegie Mellon University \cite{43}, CovidCast is another national level dashboard that shows a Mapbox level overview which is based on survey data. They show numerous parameters like hospitalized cases, percentage of the population with symptoms, etc. This data doesn’t include the statistics for individual tests but has a view of the percentage of the population(corresponding to the survey) who have been tested and are positive/negative, based on antigen tests. However, the point that needs to be taken into account here is that this dashboard is based on survey data and not validated by the government or any other facility. Similarly, governments worldwide have shared public dashboards for effective analysis of the situation. The Canadian government website \cite{50} can be used to analyze the general trend of the COVID-19 situation and shares immense knowledge right awareness resources to symptoms and treatments. Radio Canada \cite{51} is another dashboard that provides a nationwide view of Canada in terms of COVID-19 along with the testing figures. The Government of India’s Ministry of Health and Family Welfare’s website \cite{53} also provides nationwide data including daily active cases, discharged cases, deaths along with other information. New Zealand's Ministry of Health \cite{54} also displays data related to the confirmed cases, a brief analysis of the current scenario, location wise distribution of cases, the tests data and much more. Etalab \cite{55} offers a consolidated view of the official data provided by Public Health France for understanding the COVID-19 pandemic better in France. Furthermore, numerous dashboards have been developed at the local and district level to monitor community trends.

\textbf{Global level}: Since the epidemic of COVID-19, the John Hopkins dashboard has demonstrated clear and real-time updated information on the coronavirus. This dashboard offers insight into the information from countries accumulated by John Hopkins University including a rational view of the transmission of the disease from each location. This COVID-19 varies from country to country as seen on the dashboard and thus it helps us to compare trends across countries.  Instances, deaths, and recoveries are confirmed on the side panel of the John Hopkins dashboard. Automatic virus dashboard updates encourage people to identify the nature of the virus and the spread of COVID-19. The John Hopkins chart provides virus hotspots through red marks around the world on various continents. 

The Washington University dashboard gives an outline of COVID-19 infections worldwide by using confirmed cases and remaining infections to assess explanatory variables. Identifying an outbreak like COVID-19 includes the definition of an inflection point and the advantage of the public in comprehending the metamorphosis of the disease outbreak. The seafoam green slash texture on the dashboard enables users to see the movement of the pandemic and to estimate peak and curve levels. This dashboard produces information from diverse sources, including CDC and WHO, in relaying of coronavirus relevant information to the public. The simplified user design of this panel enables users to create a split second about disease outbreak by trying to compare figures. Infection, mortality, and recovery numbers on the dashboard guide users to explore the extent of the disease outbreak and to get a clear understanding of the geographical environment. The Washington University dashboard uses computational resources and technology to enhance the precision of observed trends.

Worldometer \cite{worldometer}, another powerful global dashboard goes the extra mile and shows the number of serious/critical cases, total cases per million population, deaths per million population. The dashboard also gives updated country-wise information on total tests conducted and tests per 1 million population. It gives visual representation of newly infected vs newly recovered, death rate vs recovery rate, growth factor of daily new cases and trends based on age and gender. Interestingly, it gives information on fatality rate by age, sex and comorbidity. The dashboard is very exhaustive since it provides various important graphs for all countries. The Worldometer and other dashboards immensely contribute in taking into account and understanding the depth of the COVID-19 pandemic in our area, city, region, country, and even our respective continent

Microsoft's Bing team has launched a website to follow the status of coronavirus outbreaks worldwide. The system provided all up-to-date infection statistics for each country. As an interactive map, the tracker allows people to browse the nation to see a particular number of cases and related articles from different publishers. Data are reported to be grouped from sources including WHO, the US Centers for Disease Control and Prevention (CDC), and the European Center for Disease Prevention and Control (ECDC). 

The  Coronavirus dashboard is developed by Thebaselab, and benefits by bringing a near-real-time, broad perspective of coronavirus. The color red is a tad alarmist, however, the white background is pretty well balanced. Like the JHU dashboard, The Coronavirus developed by Thebaselab \cite{57} summarises known case stats on every country which has been affected so far. Thebaselab also began publishing stories to show how the dashboard tends to work and how coronavirus compares with other major epidemics. BBC is trying to offer a good highlighter on how coronavirus has spread in the last few months. Visuals are static and the BBC also remains an important source of assessing burden of COVID-19 among the most remote and inaccessible areas globally including regions of Iran, South Korea and Italy, Comparable to the BBC, the Gray Lady's very own dashboard is doing what the New York Times does better: it gives the public an easy-to-understand education about what's going on. There are no neat looking graphics or interactive charts, but there is still a useful breaking down of how each major continent has been affected and how they are struggling to contain the virus.

\textbf{Websites related to COVID-19 testing}: Many organizations have come forward to establish an online platform to cater to COVID-19 -19 related help. While plenty of them are involved in contact tracing, there are many other websites focusing on other aspects of COVID-19 services.

These websites provide common people a user-friendly, mobile means to gather the information they are on the lookout for. For example, some websites serve the purpose of self-assessment of COVID-19 for users.  Coronavirus Self Checker \cite{58} by CDC on Microsoft Azure platform employs an interactive tool where individuals respond to COVID-19 related questions. Based on the responses, individuals are provided advice on seeking medical care. COVID-19 Screening tool \cite{59} by Apple helps identify COVID-19 related signs and suggests the next steps of action.

Some websites help users find COVID-19 test centers in their locality. Castlight came up with COVID-19 Test Site Finder \cite{60} to provide users with information on test centers in their geographical area of interest. The website also includes a self-assessment provision. A government website \cite{61} offers details about testing sites in each state. Verily has eased the process of booking appointments for COVID-19 tests through its website Project Baseline. \cite{62} 

Hence, the need for a dashboard which provides granular, structured and well formatted information without overloading the user (public, organisations, governments etc) is increasing exponentially. The dashboards can further help to understand the trends and patterns of the virus and how it has been affecting the various regions and places around the world, at the level of a university, state, nation or even global.

\section{Approaches and Challenges}
With the introduction of different solutions digital or otherwise, we are bound to come across multiple challenges that impact both the individual and community as a whole. There are several challenges, regulations, norms and solutions which govern aspects of healthcare data collection,  dissemination, analysis and its use in designing interventions or making policies which impact communities. These challenges transcend the means (physical or digital), space (geographical locations) and stakeholders (individuals, companies, governments, etc) by having effects on people who share the data, governments, healthcare providers etc. We set out to look at the different approaches currently in motion for protecting healthcare data and look at the various challenges that arise. Some of the challenges are partially addressed by existing solutions, while a large number of them still need improvisation. We examine scenarios at different socio-political levels where these challenges have unravelled differently,  followed by a brief overview of their unintended consequences.

\subsection{Current Techniques}
\subsubsection{Privacy Policies} Healthcare data in the US is protected primarily by the provisions of  Health Insurance Portability and Accountability Act, 1996 ( “HIPAA” ) which operates on a federal level. Several states and sectors have their own laws which operate in tandem with HIPAA. The strongest of them being the recent California Consumer Protection Act 2018 ( “CCPR”).\cite{63} HIPAA regulations even allow for the Preemption of State Law, wherein if a state law sets more stringent standards of protecting data and its disclosure, then the state law prevails over HIPAA. \cite{64}            

The Family Educational Rights and Privacy Act of 1974 ( “FERPA” ), The Americans with Disabilities Act of 1990 ( “ADA” ) are some examples of sector specific laws which also operate on a federal level. 

The General Data Protection Regulation ( “GDPR” ) by virtue of its extra-territorial effect \cite{65} is also relevant in this scenario since it is applicable to any entity that collects or controls the healthcare data of EU citizens in the US. 

\subsubsection{Data breach/rogue employee (security protocols)} The pie chart represents the type of breaches occurred in 2019 alone according to the information provided in \cite{66}

\begin{figure}[!h]
    \includegraphics[width = .95\linewidth]{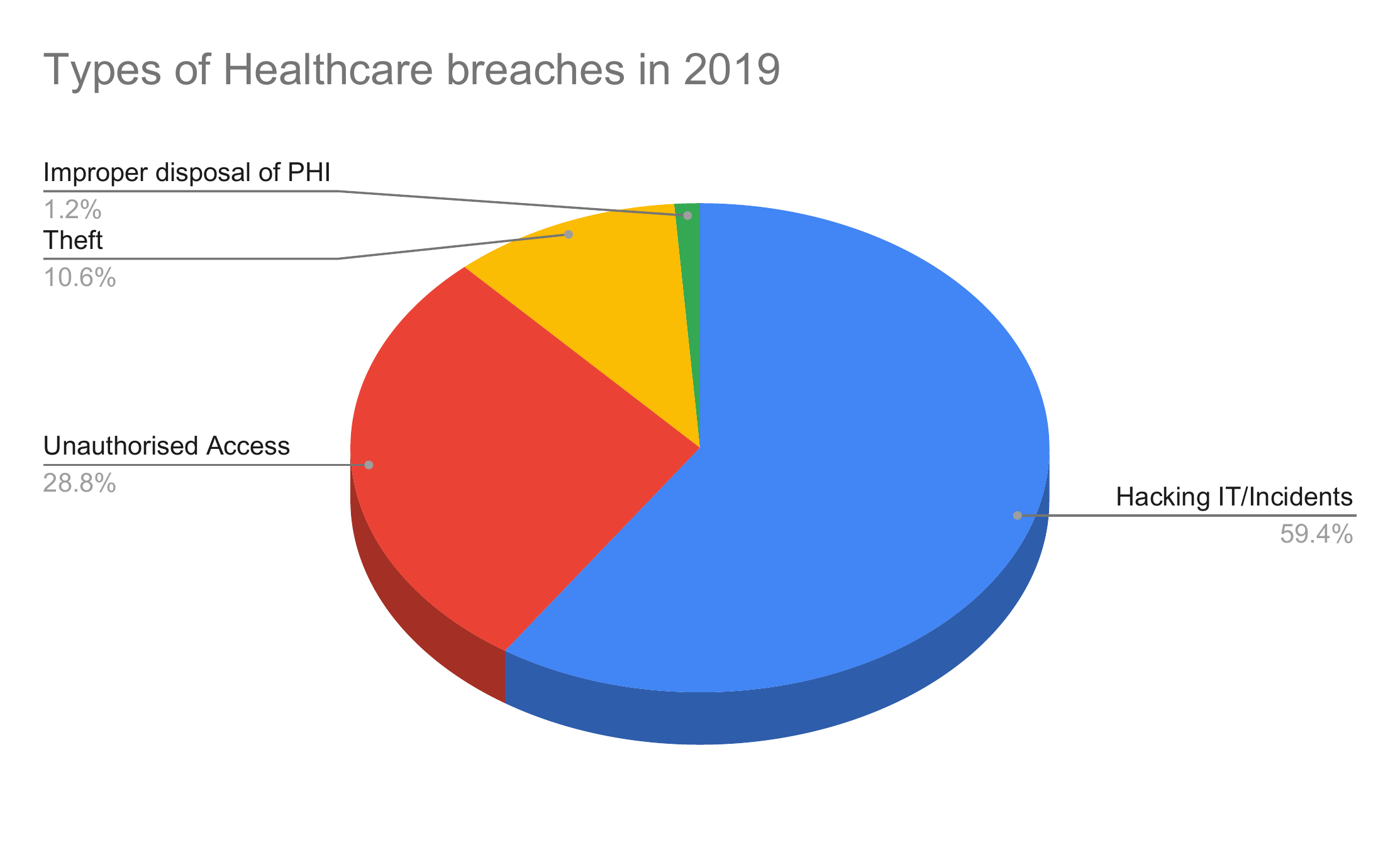}
    \centering
    \caption{Reasons for IT breaches in 2019 according to \cite{66}}
    \label{fig:pie_chart}
\end{figure}

In the past 10 years there has been an upward trend in the number of breaches which have happened. The breaches in 2019 were approximately twice more than the average number of breaches in the previous 9 years. A major cause of these breaches is owing to unauthorized access of confidential information, information technology incidents like Anthem Inc and data theft incidents, thefts etc. A brief distribution of incidents of data breach based on the type of breach is shown in the above pie chart. This is in accordance with the reports from \cite{66}. The rapid use of digital technology to manage COVID-19 data provides a unique platform for data driven loss of privacy information through multiple channels.

The unauthorized entries from employees could be due to negligence, errors, or malicious intent. Furthermore, often the loss of information happens through stolen physical records or misplaced patient records which contain a lot of PII (“\textbf{Personally Identifiable Information}”)  and PHI (“\textbf{Protected Healthcare Information}”). The fact that medical data is more life-long and cannot be changed exacerbates the issues associated, unlike in the case of any financial data breaches. This longevity makes it a long-term threat for the person whose data was compromised. Another fact to consider is that the highly interconnected nature of health care devices like - insulin pumps, pacemakers, scanning devices, etc could leave threats open to not just the leakage of information but also on the medical health of people relying on such medical aids. \cite{66}

A survey from Accenture \cite{67} in Feb 2017 revealed that 26\% of U.S. consumers or more than one in every four Americans are a victim of the healthcare data breaches. Furthermore, a large percentage of the people who took the survey (about 50\% of breach victims) suffered from medical identity theft, with an average of \$2,500 out-of-pocket costs due to the healthcare data breaches. To make matters worse, about 50\% of the respondents shared that they were not informed by the company/legal institution about the breach, i.e., they learned of the breach themselves after they'd been alerted to an error on their benefits explanation, credit card statement, or similar documents. \cite{68} These are possible pitfalls that will be accompanied by mismanagement of the large scale data collection happening in light of the COVID-19 pandemic. 

\subsubsection{Case studies} To inform our analysis we have conducted a few case-studies on the management of COVID-19 tests around the world with a special focus on South Korea, China, India, Vietnam, and Germany. Each of these countries has adopted different approaches to handling the pandemic and the sections below cover the approaches and measures taken by each of them to tackle the virus.
\begin{itemize}
    \item \textbf{India}: India remains one of the world’s hotspots for COVID-19 with rapid rise in infections (more than 9 million) \cite{69} and high recovery rates. It also provides an interesting dynamic of constitutional, intellectual and social challenges. For eg. India is very densely populated constituting several religions, and has an already volatile economy. Beyond this, India is also a big exporter of pharmaceutical and medical equipment. Further, speaking at a press briefing in Geneva, Switzerland on Monday, Mike Ryan, WHO Emergencies Programme Director, said that it was essential for India to introduce ramped up measures at a public health and societal level, to control and suppress the disease. “India is a hugely populous country”, he said. “The future of this pandemic will be determined by what happens to densely-populated countries”. \cite{70}
    
   India at the moment has no legislation that requires the protection of healthcare data and data protection in general. Within the Constitution however, the right to privacy has been found to be a fundamental right \cite{71} under the Right to Life and Liberty \cite{72}, as declared by the Supreme Court. The Information Technology Act, 2000 read with the Information Technology Rules 2011 - providing context of procedures, sensitive personal data and information reasoned with security practices - have been found to be inept. Nevertheless, there are two draft legislation: Personal Data Protection Bill,2019 (pending in Parliament) and Digital Information Security in Health Care Act (put out by the Health Ministry and specifically deals with sharing of healthcare data). These new pieces of legislation hope to bridge the gap between privacy considerations and public health.  

    On account of the pandemic, the Indian Government launched a contact tracing app “Aarogya Setu” to tackle COVID-19. The software application clearly mentions data sharing upfront in a conspicuous place when consent is gained during user registration in the Application. This helps informs the potential users about data sharing through this Application. The details of the personal information gathered, the method in which it is gathered, and by whom as well as the purpose for which it will be used is set out in its Privacy Policy \cite{73}. Besides, the Government has initiated a National Digital Health Mission as a part of which unique digital IDs will be provided as a single source of health information for the patients, government, and healthcare providers through interoperable health records. The NHA (National Health Authority) will play a crucial role in the development of these policies and managing NDHM. However, the lack of a legal framework for data protection in the country has increased the risks related to digitization.

    \textbf{Privacy issues with Aarogya setu}:
    1) App fails to clarify which ministry or department will be accessing data suggesting interdepartmental exchanges of personal information. This makes ensuring accountability challenging. 2) No legal framework to regulate the application forcing users to accept T\&C and privacy policy given by the government. 3) Uses location data via GPS trails in addition to Bluetooth, deviating from “privacy-focused global standards” which are confined to Bluetooth-based technology like TraceTogether app(Singapore) and framework suggested by MIT as a tool to monitor movement of patients and contact tracing. 4) The unique digital identity in Aarogya Setu is a static number, which increases the probability of identity breaches.
    
    \item \textbf{Vietnam}: Vietnam lacks a unified comprehensive data protection legislation. Instead, it has a very fragmented mechanism for regulating data privacy and required levels of protection through a number of laws like Civil Code, Penal Code, Law on Cyber Information Security, Law on Information Technology, Law on Cybersecurity, and sector-specific laws (“Vietnamese Data Privacy Laws”). These legal instruments, as a general principle, are personal data of individuals. While the laws fail to define the scope of personal data and the definition also varies between sectors, the laws are interpreted to protect, at a minimum, information that would enable the identification of an individual.
   The primary legislation regulating data protection is Cybersecurity law. It is to be noted that unlike cybersecurity laws in other jurisdictions (inspired by the GDPR of the EU) the Cybersecurity Law of Vietnam shares similarities with China’s Cybersecurity Law enacted in 2017. Such legal precedents focus on providing the government with the ability to control the flow of information, instead of enforcing data privacy rights for individual data subjects. \cite{74} In addition, the Law on Information Technology requires that the entity collecting personal data must provide data subjects with information about the form, scope, place, purpose of collecting, processing, and using personal data. Further, the collection, storage, use, and transfer of such data must have the consent of such a person \cite{75}.
    Vietnam’s Law on Information Technology (IT Law) requires organizations and individuals that collect, process and use personal information of other people in the network environment to obtain the consent of those people unless otherwise provided by law.Severe or Emerging Contagious Diseases fall within the definition of “Class A Infectious Diseases” under the Law on Prevention and Control of Infectious Diseases (LPCID) and individuals/ organizations in Vietnam are required to take sanitation, disinfection and sterilization measures according to instructions of competent health agencies.
    As a general principle, Vietnamese Data Privacy Laws protect information pertaining to or belonging to individuals or (to a lesser degree, organizations) that can serve to personally identify individuals (i.e., personal data). While Vietnamese Data Privacy Laws do not consistently define what information constitutes personal data and the definition also varies between sectors, the laws are interpreted to protect, at a minimum, information that would enable the identification of an individual. Specifically, the Law on Information Technology requires that the entity collecting personal data must provide data subjects with information about the form, scope, place, purpose of collecting, processing, and using personal data. Further, the collection, storage, use, and transfer of such data must have the consent of such a person.
    In the backdrop of COVID-19 , the Vietnamese gov. launched “BlueZone” (Vietnam’s Ministry of Information and Communications and Ministry of Health) \cite{76}. The following are some of the main privacy protection features of Blue Zones: \textbf{Data security}: The Bluezone app stores data locally and doesn't send any data to the system, \textbf{No location data collection}: The Bluezone app does not collect geo-location data, \textbf{Anonymity}: All Bluezone app users are anonymous to each other. Only competent and verified health authorities know those whose infected and who are suspected of infection due to close contact with COVID-19 cases, \textbf{Transparency}: The Bluezone Project codebase is completely open source.
    
    \item \textbf{South Korea}: The pandemic response of South Korea has been touted as an example to be followed by other nations. Nevertheless, its excessive testing and tracing strategy has come under the radar of privacy-related scrutiny. The data privacy in South Korea is regulated by three major legislations: Private Information Protection Act (PIPA), the Act on the Use and Protection of Credit Information (‘Credit Information Act’), and the Act on the Promotion of Information and Communications Network Utilization, these   amendments were to which were made recently. The Personal Information Protection Act (PIPA) implemented in South Korea imposes strict compliance requirements on entities that collect any potential information that can be used for identifying a specific person. Citizens also have the right to be forgotten, among other data ownership right. While organizations in the private and public sectors are legally obliged to comply with the PIPA, government agencies that require personal data for public interest purposes can collect and use data without the need to obtain consent. \cite{77} Although it is expected that organizations will use personal information on justifiable grounds by evaluating the ‘reasonable relevance’ of personal data and information that they intend to use and maintaining and preserving relevant records. Countries that sought to follow this model can face challenges by their own domestic data privacy laws especially those subjected to GDPR.The mobile application South Korea is using monitors and tracks the location of all visitors to the country, as well.  Individuals, including tourists, wear a location-tracking bracelet which is referred to as “smart city tech.”  The smart city tech platform shares information between cities on various things including traffic and pollution to find vulnerabilities and congestion, which makes it easier to identify hot spots for COVID-19. \cite{privacy_southkorea} The PIPA required data to be deleted post its utilization for the purpose it is collected.But the government has admitted that it will be permanently storing data of patients from the previous epidemic.
    \item \textbf{Germany}: 
    The German “Datenschutzkonferenz” which is a collective body composed of independent federal and state data protection authorities notified guidelines regarding COVID-19 and data protection. According to guidelines, sharing personal information of individuals infected with COVID-19 or those suspected of it would be lawful provided revealing their identity is exceptionally crucial for protecting people they were in contact with. Also, in such cases, reference should be taken to article 6(1)9c) or (f) of GDPR.For other instances, health data must be kept confidential and only be used for the purpose it was collected in the first place.Consent of data subject must be taken.
    To keep track of the number of lab tests regarding SARS-COV-2 carried out per week in Germany and how many among them are positive and negative , the RKI has initated a country-wide laboratory query in Germany. However, the quantity of laboratories reporting data seems to vary week to week. And by subtracting each change (per week)  from the cumulative total, we retrospectively work out the cumulative totals at the end of each week. Because labs are able to post-check the tests of previous calendar weeks in the test number query of RKI, the  initial figures may be revised upwards to a slight extent in subsequent reports. These sources explicitly state that the figures refer to tests conducted and that this wouldn't eqaute to the number of people tested, because of multiple tests can be done on one person. \cite{78}
    
    Germany launched an official coronavirus tracing app in June 2020 which is touted to be so secure that even government ministers can use it, though developers acknowledged few discrepancies. Germany like other EU nations have chosen against centrally storing data through its tracing app. \cite{79} This step was taken in compliance with data privacy standards. \cite{emerging_covid19} Smartphone apps have been claimed as a high-tech tools for tracking down potential COVID-19 infections. Experts state that finding new cases quickly is key to clamping down fresh clusters of cases, especially as countries progressively emerge from lockdowns and try to avoid a second wave of deaths and infections. But governments and legislative authorities in Europe have run into hurdles of legal and cultural restrictions and are trying to reconcile the necessity for effective contact tracing with the continent’s strict data privacy and digital security standards. Germany - where a person's right to their own data even after death is rooted in the constitution - has proven to be  a particular challenge. Early government suggestions to use cell tower information and GPS coordinates for the app prompted a swift backlash.

    \item \textbf{China}: Several legislations are in shape in the Republic of China for personal data protection such as National security law of the people’s republic of china, General provision of the civil law of the people’s republic of china, and cybersecurity law which in some way similar to General Data Protection Regulations by European Union. In addition, China has also introduced a draft data protection legislation recently. The CAC (Cyberspace Administration of China) released a notice on the protection of personal information when using big data joint support and defense \cite{80}. The prime objective of this notice was to set some ground rules for the use of personal information of the users to control the spread of COVID-19. Specifically, this notice addresses \textbf{Limiting Entities}: Multiple entities dealing with personal data also raises privacy issues. Only entities authorized by the National Health Commission, in accordance with cybersecurity law, should be allowed to collect personal information for public health purposes, \textbf{Data collected should have a targeted approach}: Personal information which is collected for the purpose of COVID-19 tracing should not be used for any other purpose, \textbf{Data minimization}: Limited amount of data should be collected to prevent any further breach of privacy or misuse of that data collected, and \textbf{Cybersecurity requirements}: A strict and effective implementation of cybersecurity law will help to prevent personal information from being misused.

From the above case-studies it is more than evident that we need different strategies to curb this virus across the globe, to ensure the citizens of the respective country stay safe and protected. Also, we can see the different measures taken to address the concerns of data privacy and protection, risk profiling analysis, social and racial discrimiation and economic development. These measures could help inform the decisions of countries or regions in similar situations.

\end{itemize}

\subsection{Data silos: Missed opportunity: Due to anonymization, not possible to conduct detailed analysis}

Arising from anonymization efforts in medical centers is the existence of data silos, a collection of data that is accessible by only one group or organization. As a solution for data privacy of individuals testing at medical centers, data silos may be effective, yet they bring about additional problems. Data silos prevent a possibility to conduct detailed analysis of COVID-19 testing data, which would be effective in multiple ways, including evaluating the efficacy of testing methods in different demographics. Such analysis is made impossible by data silos’ insularity, preventing health officials from seeing the bigger picture. \cite{data_silo}\cite{data_silo_2} A solution, as part of our wishlist, is to create a data lake instead; a centralized data repository optimized for analysis. Data silos are by nature decentralized, which makes it difficult to enforce a uniform protocol for data governance. A data lake fixes this additional problem due to its centralized nature. \cite{data_lake} Rather than achieving anonymization through insularity, a more intelligent approach can be taken by protecting data privacy with fine-grained, data-centric control. Furthermore, one of the primary motives of sharing the data is to learn representations and draw insights from it. Privacy preserving machine learning has been on the rise in recent times. \cite{vepakomma2018peek} Strategies like federated learning - a machine learning setting in which a centralized model is brought to localized data for training rather than data being brought to the model - can be used to ensure no explicit sharing of raw data. Split learning may also be used for this purpose; split learning is a specialized neural network method in which a designated layer in a neural network, the cut layer, ‘splits’ the network between client and server. These privacy solutions, among others, support data lakes as the best of both worlds in efficacy and privacy.
\section{Solutions that overcome privacy, mismanagement and miscommunication}        

Many governments around the world have suggested a scheme of health passports/immunity passports since a clear solution for COVID-19 management including vaccine distribution remains in its early stages and will take months before it reaches the common man globally. It is proposed that the identification of antibodies in COVID-19 virus-causing persons, the SARS-CoV-2 virus, may be used as the basis for health/immunity passports globally \cite{19}. Owing to the transient nature in terms of presence of antibodies, looking at the presence/absence of it alone could lead to issues. It might be a better option to look into vaccine based immunisation as an indicator. Adding on, we believe that a random number index is the best way to connect a user to their data, rather than a name, phone number, or email address. Later, the user will use a random number to validate their test results. The research needs to be done to understand the participant’s knowledge about various aspects such as the use of masks, the use of hand sanitizer, avoiding going to school, and college for work can prevent the spread of viruses. There are a lot of upcoming strategies and areas in which scientists and researchers are trying to explore and venture to track and avoid the exponential spread of the virus \cite{82}. These methods help to improve performance, shorten the timeline for obtaining outcomes, reduce costs and reduce the complexity involved in carrying out these studies. Some of the prominent references to current and upcoming test activities are STOPCovid \cite{STOPCovid} an initiative by MIT researchers and scientists to provide test facilities at a cheaper cost to non-laboratory experts to perform experiments in their home or private premises. Next, leveraging AI for testing, where simpler forms could be used to detect COVID, such as cough \cite{83}, breadth, and saliva. The use of these facilities will not only make the process simpler and easier to carry out, but will also allow the authorities to expand the reach of the testing and in turn, help control the available capital and supplies, but will also extend the testing to the community level, including mass population testing.

A few of the most important points to be addressed while developing the solutions are as follows: 

\begin{enumerate}
    \item \textbf{Consent of the individual}:  It should be a practice to always obtain consent of the person whose data is being collected. Also, it is the right of the individual to know about the reason and time period for the data to be utilised. 
    \item \textbf{Avoid collecting unnecessary data}: It is also suggested to avoid collecting irrelevant information while carrying out the tests such as getting information about their location, previous health history 
    \item \textbf{Authorized use of data by health officials}: The data collected should be used by the public health authorities on a need base purpose only and who are trained for the same.
    \item \textbf{Minimize sharing}: One should ensure sharing of data is minimized and is strictly limited by drafting appropriate laws and policies. 
    \item \textbf{Decentralization}: The data should be stored in a decentralised fashion using appropriate safeguarding techniques such as encryption, decentralization or de-identification to avoid or prevent any attempts to hack the data.
    \item \textbf{Training of the staff}: The staff of the organization should be trained to take the precautionary measures for containing the virus such as checking the temperature of the employees, check if they have worn masks and are maintaining a social distance, maintaining the hygiene of the food and utensils being used. 
\end{enumerate}
\section{Conclusion}
The explosive growth of diagnostic technology is a necessary component of the Strategic Emergency Preparedness Plan for Epidemic. Subsequently, the technological landscape of the development of the diagnosis of COVID-19 is rapidly coming up with new knowledge that is  formed on a daily basis. Numerous platforms for Open and rapid sharing of data has contributed to this rapid diagnostic development. In this early draft we have introduced the digital landscape of the testing workflows, the user centric issues and the challenges and consequences associated with each of these. Further we have enumerated multiple major dashboards contributing to the surveillance of COVID-19 and case studies pertinent to the digital management of resources from different countries. We have mapped the landscape of current testing systems, challenges and potential solutions. 

% \input{content/logistics}
% \input{content/health_outcomes}
% \input{content/user_centric_issues}
% \input{content/communication}
% \input{content/conclusion}

%%%%%%%%%%%%%%%%%%%%%%%%%%%%%%%%%%%%%%%%%%%%%%
%%                                          %%
%% Backmatter begins here                   %%
%%                                          %%
%%%%%%%%%%%%%%%%%%%%%%%%%%%%%%%%%%%%%%%%%%%%%%

\begin{backmatter}

\section*{Competing interests}
  The authors declare that they have no competing interests.

\section*{Acknowledgements}
  We are grateful to Riyanka Roy Choudhury, CodeX Fellow, Stanford University, Adam Berrey, CEO of PathCheck Foundation, Dr. Brooke Struck, Research Director at The Decision Lab, Canada, Vinay Gidwaney, Entrepreneur and Advisor, PathCheck Foundation, and Paola Heudebert, co-founder of Blockchain for Human Rights, Alison Tinker, Saswati Soumya, Sunny Manduva, Bhavya Pandey, and Aarathi Prasad for their assistance in discussions, support and guidance in writing of this paper.
%%  After submission of the .TEX file,                     %%
%%  you will be prompted to submit your .BBL file.         %%

% if your bibliography is in bibtex format, use those commands:

 % Style BST file (bmc-mathphys, vancouver, spbasic).
     % Bibliography file (usually '*.bib' )
\bibliography{refs}


\begin{thebibliography}{95}
% BibTex style file: bmc-mathphys.bst (version 2.0), 2013-07-15
\ifx \bisbn   \undefined \def \bisbn  #1{ISBN #1}\fi
\ifx \binits  \undefined \def \binits#1{#1}\fi
\ifx \bauthor  \undefined \def \bauthor#1{#1}\fi
\ifx \batitle  \undefined \def \batitle#1{#1}\fi
\ifx \bjtitle  \undefined \def \bjtitle#1{#1}\fi
\ifx \bvolume  \undefined \def \bvolume#1{\textbf{#1}}\fi
\ifx \byear  \undefined \def \byear#1{#1}\fi
\ifx \bissue  \undefined \def \bissue#1{#1}\fi
\ifx \bfpage  \undefined \def \bfpage#1{#1}\fi
\ifx \blpage  \undefined \def \blpage #1{#1}\fi
\ifx \burl  \undefined \def \burl#1{\textsf{#1}}\fi
\ifx \doiurl  \undefined \def \doiurl#1{\textsf{#1}}\fi
\ifx \betal  \undefined \def \betal{\textit{et al.}}\fi
\ifx \binstitute  \undefined \def \binstitute#1{#1}\fi
\ifx \binstitutionaled  \undefined \def \binstitutionaled#1{#1}\fi
\ifx \bctitle  \undefined \def \bctitle#1{#1}\fi
\ifx \beditor  \undefined \def \beditor#1{#1}\fi
\ifx \bpublisher  \undefined \def \bpublisher#1{#1}\fi
\ifx \bbtitle  \undefined \def \bbtitle#1{#1}\fi
\ifx \bedition  \undefined \def \bedition#1{#1}\fi
\ifx \bseriesno  \undefined \def \bseriesno#1{#1}\fi
\ifx \blocation  \undefined \def \blocation#1{#1}\fi
\ifx \bsertitle  \undefined \def \bsertitle#1{#1}\fi
\ifx \bsnm \undefined \def \bsnm#1{#1}\fi
\ifx \bsuffix \undefined \def \bsuffix#1{#1}\fi
\ifx \bparticle \undefined \def \bparticle#1{#1}\fi
\ifx \barticle \undefined \def \barticle#1{#1}\fi
\ifx \bconfdate \undefined \def \bconfdate #1{#1}\fi
\ifx \botherref \undefined \def \botherref #1{#1}\fi
\ifx \url \undefined \def \url#1{\textsf{#1}}\fi
\ifx \bchapter \undefined \def \bchapter#1{#1}\fi
\ifx \bbook \undefined \def \bbook#1{#1}\fi
\ifx \bcomment \undefined \def \bcomment#1{#1}\fi
\ifx \oauthor \undefined \def \oauthor#1{#1}\fi
\ifx \citeauthoryear \undefined \def \citeauthoryear#1{#1}\fi
\ifx \endbibitem  \undefined \def \endbibitem {}\fi
\ifx \bconflocation  \undefined \def \bconflocation#1{#1}\fi
\ifx \arxivurl  \undefined \def \arxivurl#1{\textsf{#1}}\fi
\csname PreBibitemsHook\endcsname

%%% 1
\bibitem{1}
\begin{barticle}
\bauthor{\bsnm{Aguiar}, \binits{E.R.G.R.}},
\bauthor{\bsnm{Navas}, \binits{J.}},
\bauthor{\bsnm{Pacheco}, \binits{L.G.C.}}:
\batitle{The covid-19 diagnostic technology landscape: Efficient data sharing
  drives diagnostic development}.
\bjtitle{Frontiers in Public Health}
\bvolume{8},
\bfpage{309}
(\byear{2020}).
doi:\doiurl{10.3389/fpubh.2020.00309}
\end{barticle}
\endbibitem

%%% 2
\bibitem{2}
\begin{barticle}
\bauthor{\bsnm{Moorthy}, \binits{V.}},
\bauthor{\bsnm{Henao~Restrepo}, \binits{A.M.}},
\bauthor{\bsnm{Preziosi}, \binits{M.-P.}},
\bauthor{\bsnm{Swaminathan}, \binits{S.}}:
\batitle{Data sharing for novel coronavirus (covid-19)}.
\bjtitle{Bulletin of the World Health Organization}
\bvolume{98}(\bissue{3}),
\bfpage{150}--\blpage{150}
(\byear{2020-3-01})
\end{barticle}
\endbibitem

%%% 3
\bibitem{3}
\begin{botherref}
\oauthor{\bsnm{Cosgrove}, \binits{C.}}:
Privacy questions for COVID-19 testing and health monitoring
(2020).
\url{https://iapp.org/news/a/privacy-questions-for-covid-19-testing-and-health-monitoring/}
\end{botherref}
\endbibitem

%%% 4
\bibitem{4}
\begin{botherref}
\oauthor{\bsnm{Ahmadi}, \binits{S.}},
\oauthor{\bsnm{Bempong}, \binits{N.-E.}},
\oauthor{\bsnm{Santis}, \binits{O.D.}},
\oauthor{\bsnm{Sheath}, \binits{D.}},
\oauthor{\bsnm{Flahault}, \binits{A.}}:
The role of digital technologies in tackling the zika outbreak: a scoping
  review.
Journal of Public Health and Emergency
\textbf{2}(6)
(2018)
\end{botherref}
\endbibitem

%%% 5
\bibitem{5}
\begin{botherref}
\oauthor{\bsnm{Volbrecht}, \binits{A.}}:
The promise of a digitally connected DR Congo
(2019).
\url{https://www.path.org/articles/digital-congo-ebola/}
\end{botherref}
\endbibitem

%%% 6
\bibitem{6}
\begin{botherref}
\oauthor{\bsnm{Bempong}, \binits{N.-E.}},
\oauthor{\bparticle{Ruiz~de} \bsnm{Castaneda}, \binits{R.}},
\oauthor{\bsnm{Schütte}, \binits{S.}},
\oauthor{\bsnm{Bolon}, \binits{I.}},
\oauthor{\bsnm{Keiser}, \binits{O.}},
\oauthor{\bsnm{Escher}, \binits{G.}},
\oauthor{\bsnm{Flahault}, \binits{A.}}:
Precision global health - the case of ebola: A scoping review.
Journal of Global Health
\textbf{9}
(2019).
doi:\doiurl{10.7189/jogh.09.010404}
\end{botherref}
\endbibitem

%%% 7
\bibitem{7}
\begin{barticle}
\bauthor{\bsnm{Tambo}, \binits{E.}},
\bauthor{\bsnm{Adama}, \binits{K.}},
\bauthor{\bsnm{Talla}, \binits{M.}},
\bauthor{\bsnm{CF}, \binits{C.}},
\bauthor{\bsnm{C.}, \binits{F.}}:
\batitle{Digital technology and mobile applications impact on zika and ebola
  epidemics data sharing and emergency response}.
\bjtitle{J Health Med Informatics}
\bvolume{8},
\bfpage{254}
(\byear{2017}).
doi:\doiurl{10.4172/2157-7420.1000254}
\end{barticle}
\endbibitem

%%% 8
\bibitem{8}
\begin{botherref}
\oauthor{\bsnm{Reiny}, \binits{S.}}:
Using NASA Satellite Data to Predict Malaria Outbreaks
(2017).
\url{https://www.nasa.gov/feature/goddard/2017/using-nasa-satellite-data-to-predict-malaria-outbreaks}
\end{botherref}
\endbibitem

%%% 9
\bibitem{9}
\begin{botherref}
Digital solutions for malaria elimination community of practice
(2017)
\end{botherref}
\endbibitem

%%% 10
\bibitem{cdcflowscreening}
\begin{botherref}
Example of layout and flow of individuals being screened
(2020).
\url{https://www.cdc.gov/coronavirus/2019-ncov/images/hcp/figure-layout-flow-screening.png}
\end{botherref}
\endbibitem

%%% 11
\bibitem{10}
\begin{botherref}
Performing Broad-Based Testing for SARS-CoV-2 in Congregate Settings
(2020).
\url{https://www.cdc.gov/coronavirus/2019-ncov/hcp/broad-based-testing.html}
\end{botherref}
\endbibitem

%%% 12
\bibitem{11}
\begin{botherref}
NHS Testing and tracing for coronavirus
(2020).
\url{https://www.nhs.uk/conditions/coronavirus-covid-19/testing-and-tracing/}
\end{botherref}
\endbibitem

%%% 13
\bibitem{12}
\begin{botherref}
MIT Medical FAQ Testing for COVID-19
(2020).
\url{https://medical.mit.edu/faqs/faq-testing-covid-19}
\end{botherref}
\endbibitem

%%% 14
\bibitem{13}
\begin{botherref}
Minnesota Community Testing
(2020).
\url{https://www.health.state.mn.us/diseases/coronavirus/testsites/community.html}
\end{botherref}
\endbibitem

%%% 15
\bibitem{14}
\begin{botherref}
Henry Ford COVID-19 Symptoms Testing
(2020).
\url{https://www.henryford.com/coronavirus/covid19-symptoms-testing-treatment}
\end{botherref}
\endbibitem

%%% 16
\bibitem{15}
\begin{botherref}
CareWell COVID-19 Testing
(2020).
\url{https://www.carewellurgentcare.com/covid-19-response-at-carewell/}
\end{botherref}
\endbibitem

%%% 17
\bibitem{19}
\begin{barticle}
\bauthor{\bsnm{Brown}, \binits{B.}},
\bauthor{\bsnm{Kelly}, \binits{D.}},
\bauthor{\bsnm{Wilkinson}, \binits{D.}},
\bauthor{\bsnm{Savulescu}, \binits{J.}}:
\batitle{The scientific and ethical feasibility of immunity passports}.
\bjtitle{The Lancet Infectious Diseases}
(\byear{2020}).
doi:\doiurl{10.1016/S1473-3099(20)30766-0}
\end{barticle}
\endbibitem

%%% 18
\bibitem{22}
\begin{botherref}
\oauthor{\bsnm{Olivarius}, \binits{K.}}:
The Dangerous History of Immunoprivilege
(2020).
\url{https://www.nytimes.com/2020/04/12/opinion/coronavirus-immunity-passports.html#click=https://t.co/QcXDROj5IL}
\end{botherref}
\endbibitem

%%% 19
\bibitem{23}
\begin{barticle}
\bauthor{\bsnm{Kind}, \binits{C.}}:
\batitle{Exit through the app store?}
\bjtitle{Patterns}
\bvolume{1}(\bissue{3}),
\bfpage{100054}
(\byear{2020}).
doi:\doiurl{10.1016/j.patter.2020.100054}
\end{barticle}
\endbibitem

%%% 20
\bibitem{18}
\begin{botherref}
\oauthor{\bsnm{Gruener}, \binits{D.}}:
Immunity Certificates: If We Must Have Them, We Must Do It Right
(2020).
\url{https://ethics.harvard.edu/files/center-for-ethics/files/12immunitycertificates.pdf}
\end{botherref}
\endbibitem

%%% 21
\bibitem{16}
\begin{botherref}
Health Passport Europe
(2020).
\url{https://www.healthpassportireland.ie/}
\end{botherref}
\endbibitem

%%% 22
\bibitem{17}
\begin{botherref}
\oauthor{\bsnm{Paul~Mozur}, \binits{R.Z.}},
\oauthor{\bsnm{Krolik}, \binits{A.}}:
In Coronavirus Fight, China Gives Citizens a Color Code, With Red Flags
(2020).
\url{https://www.nytimes.com/2020/03/01/business/china-coronavirus-surveillance.html}
\end{botherref}
\endbibitem

%%% 23
\bibitem{20}
\begin{botherref}
\oauthor{\bsnm{Jr.}, \binits{B.L.}}:
WHO doesn’t recommend coronavirus passports, because immunity remains
  questionable
(2020).
\url{https://www.cnbc.com/2020/09/16/who-doesnt-recommend-coronavirus-passports-because-immunity-remains-questionable.html}
\end{botherref}
\endbibitem

%%% 24
\bibitem{21}
\begin{botherref}
\oauthor{\bsnm{WHO}}:
"Immunity passports" in the context of COVID-19
(2020).
\url{https://www.who.int/publications/i/item/immunity-passports-in-the-context-of-covid-19}
\end{botherref}
\endbibitem

%%% 25
\bibitem{24}
\begin{botherref}
IBM Digital Health Pass
(2020).
\url{https://www.ibm.com/products/digital-health-pass}
\end{botherref}
\endbibitem

%%% 26
\bibitem{25}
\begin{botherref}
V-Health: The People's Passport
(2020).
\url{https://v-healthpassport.co.uk/}
\end{botherref}
\endbibitem

%%% 27
\bibitem{26}
\begin{botherref}
WISeKey’s WIShelter Covid-19 Platform Using Digital IDs and Blockchain to
  Help Tourist Destinations Certify Travelers are not Infected
(2020).
\url{https://www.wisekey.com/press/wisekeys-wishelter-covid-19-platform-using-digital-ids-and-blockchain-to-help-tourist-destin\\ations-certify-travelers-are-not-infected/}
\end{botherref}
\endbibitem

%%% 28
\bibitem{covipass}
\begin{botherref}
COVI-Pass by Tento Health
(2020).
\url{https://tentohealth.com/}
\end{botherref}
\endbibitem

%%% 29
\bibitem{27}
\begin{botherref}
\oauthor{\bsnm{SecureIdNews}}:
Digital ID immunity passports gain steam around the globe
(2020).
\url{https://www.secureidnews.com/news-item/digital-id-immunity-passports-gain-steam-around-the-globe/}
\end{botherref}
\endbibitem

%%% 30
\bibitem{28}
\begin{botherref}
International monitor: public health identity systems, Ada Lovelace Institute
(2020).
\url{https://www.adalovelaceinstitute.org/project/international-monitor-public-health-identity-systems/}
\end{botherref}
\endbibitem

%%% 31
\bibitem{29}
\begin{botherref}
\oauthor{\bsnm{PRNewswire}}:
Irish-based ROQU Group launches world-first 'Health Passport' digital platform
  to support increased global COVID-19 testing
(2020).
\url{https://www.prnewswire.com/news-releases/irish-based-roqu-group-launches-world-first-health-passport-digital-platform-to-sup\\port-increased-global-covid-19-testing-301120037.html}
\end{botherref}
\endbibitem

%%% 32
\bibitem{30}
\begin{botherref}
ALHOSN: Protecting yourself protects your community
(2020).
\url{https://www.alhosnapp.ae/en/home/}
\end{botherref}
\endbibitem

%%% 33
\bibitem{31}
\begin{botherref}
WHO Digital COVID-19 Vaccination Infrastructure
(2020).
\url{https://guardtime.com/blog/who-digital-covid-19-vaccination-infrastructure}
\end{botherref}
\endbibitem

%%% 34
\bibitem{32}
\begin{botherref}
COVID-19 Credentials Initiative
(2020).
\url{https://www.covidcreds.com/}
\end{botherref}
\endbibitem

%%% 35
\bibitem{United_and_cathay}
\begin{botherref}
\oauthor{\bsnm{SMITS}, \binits{J.}}:
Digital health passports roll out for air travelers tracking COVID status,
  vaccination proof
(2020).
\url{https://www.lifesitenews.com/blogs/digital-health-passports-roll-out-for-air-travelers-tracking-covid-status-vaccination-proof}
\end{botherref}
\endbibitem

%%% 36
\bibitem{common_pass}
\begin{botherref}
CommonPass by The Commons Project
(2020).
\url{https://thecommonsproject.org/commonpass}
\end{botherref}
\endbibitem

%%% 37
\bibitem{common_pass_site}
\begin{botherref}
CommonPass
(2020).
\url{https://commonpass.org/}
\end{botherref}
\endbibitem

%%% 38
\bibitem{health_pass}
\begin{botherref}
Health Pass by CLEAR
(2020).
\url{https://www.clearme.com/healthpass}
\end{botherref}
\endbibitem

%%% 39
\bibitem{hi_card}
\begin{botherref}
Hi+ Card by TDDS
(2020).
\url{https://hicard.travel/}
\end{botherref}
\endbibitem

%%% 40
\bibitem{33}
\begin{botherref}
\oauthor{\bsnm{Gandhi}, \binits{D.}},
\oauthor{\bsnm{Landage}, \binits{S.}},
\oauthor{\bsnm{Bae}, \binits{J.}},
\oauthor{\bsnm{Shankar}, \binits{S.}},
\oauthor{\bsnm{Sukumaran}, \binits{R.}},
\oauthor{\bsnm{Patwa}, \binits{P.}},
\oauthor{\bsnm{au2}, \binits{S.T.V.}},
\oauthor{\bsnm{Katiyar}, \binits{P.}},
\oauthor{\bsnm{Advani}, \binits{S.}},
\oauthor{\bsnm{Iyer}, \binits{R.}},
\oauthor{\bsnm{Anand}, \binits{S.}},
\oauthor{\bsnm{Mahindra}, \binits{A.}},
\oauthor{\bsnm{Barbar}, \binits{R.}},
\oauthor{\bsnm{Singh}, \binits{A.}},
\oauthor{\bsnm{Raskar}, \binits{R.}}:
Clinical Landscape of COVID-19 Testing: Difficult Choices
(2020).
\arxivurl{2011.04202}
\end{botherref}
\endbibitem

%%% 41
\bibitem{34}
\begin{botherref}
Clinical Laboratory Improvement Amendments (CLIA)
(2020).
\url{https://www.cdc.gov/clia/index.html}
\end{botherref}
\endbibitem

%%% 42
\bibitem{35}
\begin{botherref}
Quest Diagnostics, COVID-19
(2020).
\url{https://www.questdiagnostics.com/home/Covid-19/}
\end{botherref}
\endbibitem

%%% 43
\bibitem{36}
\begin{botherref}
COVID-19 Laboratory Testing: Medpace Central Labs
(2020).
\url{https://www.medpace.com/covid-19-laboratory-testing/}
\end{botherref}
\endbibitem

%%% 44
\bibitem{37}
\begin{botherref}
Flexible, agile and proactive laboratory services dedicated to global clinical
  development
(2020).
\url{https://www.iconplc.com/services/laboratories/}
\end{botherref}
\endbibitem

%%% 45
\bibitem{38}
\begin{botherref}
Syneos Health - Collaborating to Address the Challenge of COVID-19
(2020).
\url{https://www.syneoshealth.com/covid-19-resource-center}
\end{botherref}
\endbibitem

%%% 46
\bibitem{39}
\begin{botherref}
Verily: Delivering COVID-19 screening and testing through Project Baseline and
  community-based sites
(2020).
\url{https://verily.com/solutions/covid-19-testing/}
\end{botherref}
\endbibitem

%%% 47
\bibitem{41}
\begin{botherref}
2019 Novel Coronavirus Disease (COVID-19)
(2020).
\url{https://www.bioreference.com/coronavirus/}
\end{botherref}
\endbibitem

%%% 48
\bibitem{42}
\begin{botherref}
Find Diseases \& Conditions
(2020).
\url{https://www.mayoclinic.org}
\end{botherref}
\endbibitem

%%% 49
\bibitem{44}
\begin{botherref}
Amherst College COVID-19 Dashboard
(2020).
\url{https://www.amherst.edu/news/covid-19/dashboard}
\end{botherref}
\endbibitem

%%% 50
\bibitem{dashboards}
\begin{botherref}
We Rate COVID-19 Dashboards
(2020).
\url{https://www.ratecoviddashboard.com/}
\end{botherref}
\endbibitem

%%% 51
\bibitem{45}
\begin{botherref}
Wagner College COVID-19 Dashboard
(2020).
\url{https://sites.google.com/wagner.edu/covid-19casetracker}
\end{botherref}
\endbibitem

%%% 52
\bibitem{46}
\begin{botherref}
Tulane University COVID-19 Dashboard
(2020).
\url{https://tulane.edu/covid-19/dashboard}
\end{botherref}
\endbibitem

%%% 53
\bibitem{47}
\begin{botherref}
Ohio State University COVID-19 Dashboard
(2020).
\url{https://news.osu.edu/a-deeper-dive-into-ohio-states-top-rated-covid-19-testing-data-dashboard/}
\end{botherref}
\endbibitem

%%% 54
\bibitem{48}
\begin{botherref}
George Mason University COVID-19 Dashboard
(2020).
\url{https://www2.gmu.edu/campus-covid-19-data}
\end{botherref}
\endbibitem

%%% 55
\bibitem{49}
\begin{botherref}
CDC COVID Data Tracker
(2020).
\url{https://covid.cdc.gov/covid-data-tracker/?CDC_AA_refVal=https%3A%2F%2Fwww.cdc.gov%2Fcoronavirus%2F2019-ncov%2Fcases-updates%2Fcases-in-us.html#cases_casesper100klast7days}
\end{botherref}
\endbibitem

%%% 56
\bibitem{43}
\begin{botherref}
Carnegie Mellon University COVID-19 Dashboard
(2020).
\url{https://covidcast.cmu.edu/?sensor=doctor-visits-smoothed_adj_cli&level=county&date=20201027&signalType=value&encoding=color&mode=single&region=42003}
\end{botherref}
\endbibitem

%%% 57
\bibitem{50}
\begin{botherref}
Coronavirus disease (COVID-19): Outbreak update, Government of Canada
(2020).
\url{https://www.canada.ca/en/public-health/services/diseases/2019-novel-coronavirus-infection.html?topic=tilelink}
\end{botherref}
\endbibitem

%%% 58
\bibitem{51}
\begin{botherref}
Evolution of COVID-19, CBC Radio-Canada
(2020).
\url{https://ici.radio-canada.ca/info/2020/coronavirus-covid-19-pandemie-cas-carte-maladie-symptomes-propagation/index-en.html}
\end{botherref}
\endbibitem

%%% 59
\bibitem{53}
\begin{botherref}
COVID-19 State wise Status, Ministry of Health and Family Welfare, India
(2020).
\url{https://www.mohfw.gov.in/}
\end{botherref}
\endbibitem

%%% 60
\bibitem{54}
\begin{botherref}
COVID-19 : Current cases, New Zealand Ministry of Health
(2020).
\url{https://www.health.govt.nz/our-work/diseases-and-conditions/covid-19-novel-coronavirus/covid-19-data-and-statistics/covid-19-current-cases}
\end{botherref}
\endbibitem

%%% 61
\bibitem{55}
\begin{botherref}
France COVID-19 Dashboard
(2020).
\url{https://dashboard.covid19.data.gouv.fr/vue-d-ensemble?location=FRA}
\end{botherref}
\endbibitem

%%% 62
\bibitem{worldometer}
\begin{botherref}
Worldometer
(2020).
\url{https://www.worldometers.info/coronavirus/}
\end{botherref}
\endbibitem

%%% 63
\bibitem{57}
\begin{botherref}
thebaselab
(2020).
\url{https://coronavirus.thebaselab.com}
\end{botherref}
\endbibitem

%%% 64
\bibitem{58}
\begin{botherref}
CDC COVID-19 Testing Overview
(2020).
\url{https://www.cdc.gov/coronavirus/2019-ncov/symptoms-testing/testing.html}
\end{botherref}
\endbibitem

%%% 65
\bibitem{59}
\begin{botherref}
Apple COVID-19 Screening Tool
(2020).
\url{https://covid19.apple.com/screening#donotlink}
\end{botherref}
\endbibitem

%%% 66
\bibitem{60}
\begin{botherref}
Castlight Inc. COVID-19 Test Site Finder: Get Information on Coronavirus
  Testing Near You
(2020).
\url{https://my.castlighthealth.com/corona-virus-testing-sites/}
\end{botherref}
\endbibitem

%%% 67
\bibitem{61}
\begin{botherref}
Community-Based Testing Sites for COVID-19
(2020).
\url{https://www.hhs.gov/coronavirus/community-based-testing-sites/index.html}
\end{botherref}
\endbibitem

%%% 68
\bibitem{62}
\begin{botherref}
Project Baseline by verily
(2020).
\url{https://www.projectbaseline.com/study/covid-19/}
\end{botherref}
\endbibitem

%%% 69
\bibitem{63}
\begin{botherref}
Comparing privacy laws: GDPR v. CCPA
(2020).
\url{https://fpf.org/wp-content/uploads/2018/11/GDPR_CCPA_Comparison-Guide.pdf}
\end{botherref}
\endbibitem

%%% 70
\bibitem{64}
\begin{botherref}
Deliberation on HIPAA Privacy Rule
(2020).
\url{https://www.hhs.gov/hipaa/for-professionals/faq/405/is-this-more-protective-state-law-preempted-by-the-privacy-rule/index.html}
\end{botherref}
\endbibitem

%%% 71
\bibitem{65}
\begin{botherref}
Does the GDPR apply to companies outside of the EU?
(2020).
\url{https://gdpr.eu/companies-outside-of-europe}
\end{botherref}
\endbibitem

%%% 72
\bibitem{66}
\begin{botherref}
Healthcare Data Breach Statistics, HIPAA Journal
(2020).
\url{https://www.hipaajournal.com/healthcare-data-breach-statistics/}
\end{botherref}
\endbibitem

%%% 73
\bibitem{67}
\begin{botherref}
One in Four US Consumers Have Had Their Healthcare Data Breached, Accenture
  Survey Reveals
(2020).
\url{https://newsroom.accenture.com/subjects/technology/one-in-four-us-consumers-have-had-their-healthcare-data-breached-accenture-survey-reveals.html}
\end{botherref}
\endbibitem

%%% 74
\bibitem{68}
\begin{botherref}
Top 10 Biggest Healthcare Data Breaches of All Time
(2020).
\url{https://digitalguardian.com/blog/top-10-biggest-healthcare-data-breaches-all-time}
\end{botherref}
\endbibitem

%%% 75
\bibitem{69}
\begin{botherref}
John Hopkins University Mortality Analyses
(2020).
\url{https://coronavirus.jhu.edu/data/mortality}
\end{botherref}
\endbibitem

%%% 76
\bibitem{70}
\begin{botherref}
COVID-19: Lockdown across India, in line with WHO guidance
(2020).
\url{https://news.un.org/en/story/2020/03/1060132}
\end{botherref}
\endbibitem

%%% 77
\bibitem{71}
\begin{botherref}
\oauthor{\bsnm{Grover}, \binits{A.}}:
COVID-19 in India: Lockdown, Legal Challenges, and Disparate Impacts
(2020).
\url{https://blog.petrieflom.law.harvard.edu/2020/05/18/india-global-responses-covid19/}
\end{botherref}
\endbibitem

%%% 78
\bibitem{72}
\begin{botherref}
THE CONSTITUTION OF INDIA
(2020).
\url{https://www.india.gov.in/sites/upload_files/npi/files/coi_part_full.pdf}
\end{botherref}
\endbibitem

%%% 79
\bibitem{73}
\begin{botherref}
Frequently Asked Questions on Aarogya Setu App
(2020).
\url{https://covid19.assam.gov.in/wp-content/uploads/2020/05/Aarogya_Setu_FAQ.pdf}
\end{botherref}
\endbibitem

%%% 80
\bibitem{74}
\begin{botherref}
DATA PROTECTION LAWS OF THE WORLD
(2020).
\url{https://www.dlapiperdataprotection.com/index.html?t=law&c=VN}
\end{botherref}
\endbibitem

%%% 81
\bibitem{75}
\begin{botherref}
\oauthor{\bsnm{McKenzie}, \binits{B.}}:
Data Privacy \& Security Survey
(2020).
\url{https://www.bakermckenzie.com/-/media/files/insight/publications/2020/04/covid19-data-privacy--security-survey17-april.pdf}
\end{botherref}
\endbibitem

%%% 82
\bibitem{76}
\begin{botherref}
"Protect yourself" Ministry of Health Vietnam
(2020).
\url{https://bluezone.ai}
\end{botherref}
\endbibitem

%%% 83
\bibitem{77}
\begin{botherref}
\oauthor{\bsnm{Helen~Chan}, \binits{R.I.E.}}:
Pervasive personal data collection at the heart of South Korea’s COVID-19
  success may not translate
(2020).
\url{https://blogs.thomsonreuters.com/answerson/south-korea-covid-19-data-privacy/}
\end{botherref}
\endbibitem

%%% 84
\bibitem{privacy_southkorea}
\begin{botherref}
\oauthor{\bsnm{Panakal}, \binits{D.D.}}:
Privacy vs. pandemic control in south korea
(2020)
\end{botherref}
\endbibitem

%%% 85
\bibitem{78}
\begin{botherref}
\oauthor{\bsnm{Frank~Jordans}, \binits{A.P.}}:
Coronavirus tracing app a test for privacy-minded Germany
(2020).
\url{https://abcnews.go.com/Health/wireStory/coronavirus-tracing-app-test-privacy-minded-germany-71270334}
\end{botherref}
\endbibitem

%%% 86
\bibitem{79}
\begin{botherref}
Mobile applications in support of contact tracing for COVID-19, European Centre
  for Disease Prevention and Control
(2020).
\url{https://www.ecdc.europa.eu/sites/default/files/documents/covid-19-mobile-applications-contact-tracing.pdf}
\end{botherref}
\endbibitem

%%% 87
\bibitem{emerging_covid19}
\begin{botherref}
\oauthor{\bsnm{Lothar~Wieler}, \binits{R.G.} \bsuffix{Ute~Rexroth}}:
Emerging covid-19 success story: Germany’s strong enabling environment
(2020)
\end{botherref}
\endbibitem

%%% 88
\bibitem{80}
\begin{botherref}
\oauthor{\bsnm{Carolyn~Bigg}, \binits{F.S.} \bsuffix{Venus~Cheung}}:
Navigating China Episode 14: New draft national, harmonised data protection law
  for Mainland China Navigating China: The digital journey
(2020).
\url{https://www.dlapiper.com/en/us/insights/publications/2020/10/navigating-china-episode-14/}
\end{botherref}
\endbibitem

%%% 89
\bibitem{data_silo}
\begin{botherref}
\oauthor{\bsnm{Alienor}}:
What is a Data Silo and Why is It Bad for Your Organization?
(2020).
\url{https://www.plixer.com/blog/data-silo-what-is-it-why-is-it-bad/}
\end{botherref}
\endbibitem

%%% 90
\bibitem{data_silo_2}
\begin{botherref}
What are Data Silos?
(2020).
\url{https://www.talend.com/resources/what-are-data-silos/}
\end{botherref}
\endbibitem

%%% 91
\bibitem{data_lake}
\begin{botherref}
\oauthor{\bsnm{Turajski}, \binits{N.}}:
Operationalizing Data Lake Privacy Governance for Value Creation
(2020).
\url{https://www.talend.com/resources/what-are-data-silos/}
\end{botherref}
\endbibitem

%%% 92
\bibitem{vepakomma2018peek}
\begin{botherref}
\oauthor{\bsnm{Vepakomma}, \binits{P.}},
\oauthor{\bsnm{Swedish}, \binits{T.}},
\oauthor{\bsnm{Raskar}, \binits{R.}},
\oauthor{\bsnm{Gupta}, \binits{O.}},
\oauthor{\bsnm{Dubey}, \binits{A.}}:
No Peek: A Survey of private distributed deep learning
(2018).
\arxivurl{1812.03288}
\end{botherref}
\endbibitem

%%% 93
\bibitem{82}
\begin{botherref}
\oauthor{\bsnm{Nguyen}, \binits{T.T.}}:
Artificial Intelligence in the Battle against Coronavirus (COVID-19): A Survey
  and Future Research Directions
(2020).
\url{https://www.researchgate.net/publication/340487417_Artificial_Intelligence_in_the_Battle_against_Coronavirus_COVID-19_A_Survey_and_Future_Research_Directions}
\end{botherref}
\endbibitem

%%% 94
\bibitem{STOPCovid}
\begin{barticle}
\bauthor{\bsnm{Joung}, \binits{J.}},
\bauthor{\bsnm{Ladha}, \binits{A.}},
\bauthor{\bsnm{Saito}, \binits{M.}},
\bauthor{\bsnm{Segel}, \binits{M.}},
\bauthor{\bsnm{Bruneau}, \binits{R.}},
\bauthor{\bsnm{Huang}, \binits{M.-l.W.}},
\bauthor{\bsnm{Kim}, \binits{N.-G.}},
\bauthor{\bsnm{Yu}, \binits{X.}},
\bauthor{\bsnm{Li}, \binits{J.}},
\bauthor{\bsnm{Walker}, \binits{B.D.}},
\bauthor{\bsnm{Greninger}, \binits{A.L.}},
\bauthor{\bsnm{Jerome}, \binits{K.R.}},
\bauthor{\bsnm{Gootenberg}, \binits{J.S.}},
\bauthor{\bsnm{Abudayyeh}, \binits{O.O.}},
\bauthor{\bsnm{Zhang}, \binits{F.}}:
\batitle{Point-of-care testing for covid-19 using sherlock diagnostics}.
\bjtitle{medRxiv}
(\byear{2020}).
doi:\doiurl{10.1101/2020.05.04.20091231}.
\arxivurl{https://www.medrxiv.org/content/early/2020/05/08/2020.05.04.20091231.full.pdf}
\end{barticle}
\endbibitem

%%% 95
\bibitem{83}
\begin{botherref}
\oauthor{\bsnm{Chu}, \binits{J.}}:
Artificial intelligence model detects asymptomatic Covid-19 infections through
  cellphone-recorded coughs
(2020).
\url{https://news.mit.edu/2020/covid-19-cough-cellphone-detection-1029}
\end{botherref}
\endbibitem

\end{thebibliography}

\newcommand{\BMCxmlcomment}[1]{}

\BMCxmlcomment{

<refgrp>

<bibl id="B1">
  <title><p>The COVID-19 Diagnostic Technology Landscape: Efficient Data
  Sharing Drives Diagnostic Development</p></title>
  <aug>
    <au><snm>Aguiar</snm><fnm>ERGR</fnm></au>
    <au><snm>Navas</snm><fnm>J</fnm></au>
    <au><snm>Pacheco</snm><fnm>LGC</fnm></au>
  </aug>
  <source>Frontiers in Public Health</source>
  <pubdate>2020</pubdate>
  <volume>8</volume>
  <fpage>309</fpage>
  <url>https://www.frontiersin.org/article/10.3389/fpubh.2020.00309</url>
</bibl>

<bibl id="B2">
  <title><p>Data sharing for novel coronavirus (COVID-19)</p></title>
  <aug>
    <au><snm>Moorthy</snm><fnm>V</fnm></au>
    <au><snm>Henao Restrepo</snm><fnm>AM</fnm></au>
    <au><snm>Preziosi</snm><fnm>MP</fnm></au>
    <au><snm>Swaminathan</snm><fnm>S</fnm></au>
  </aug>
  <source>Bulletin of the World Health Organization</source>
  <publisher>World Health Organization</publisher>
  <pubdate>2020-3-01</pubdate>
  <volume>98</volume>
  <issue>3</issue>
  <fpage>150</fpage>
  <lpage>150</lpage>
</bibl>

<bibl id="B3">
  <title><p>Privacy questions for COVID-19 testing and health
  monitoring</p></title>
  <aug>
    <au><snm>Cosgrove</snm><fnm>C</fnm></au>
  </aug>
  <pubdate>2020</pubdate>
  <url>https://iapp.org/news/a/privacy-questions-for-covid-19-testing-and-health-monitoring/</url>
</bibl>

<bibl id="B4">
  <title><p>The role of digital technologies in tackling the Zika outbreak: a
  scoping review</p></title>
  <aug>
    <au><snm>Ahmadi</snm><fnm>S</fnm></au>
    <au><snm>Bempong</snm><fnm>NE</fnm></au>
    <au><snm>Santis</snm><fnm>OD</fnm></au>
    <au><snm>Sheath</snm><fnm>D</fnm></au>
    <au><snm>Flahault</snm><fnm>A</fnm></au>
  </aug>
  <source>Journal of Public Health and Emergency</source>
  <pubdate>2018</pubdate>
  <volume>2</volume>
  <issue>6</issue>
  <url>http://jphe.amegroups.com/article/view/4556</url>
</bibl>

<bibl id="B5">
  <title><p>The promise of a digitally connected DR Congo</p></title>
  <aug>
    <au><snm>Volbrecht</snm><fnm>A</fnm></au>
  </aug>
  <pubdate>2019</pubdate>
  <url>https://www.path.org/articles/digital-congo-ebola/</url>
</bibl>

<bibl id="B6">
  <title><p>Precision Global Health - The case of Ebola: A scoping
  review</p></title>
  <aug>
    <au><snm>Bempong</snm><fnm>NE</fnm></au>
    <au><snm>Castaneda</snm><fnm>R</fnm></au>
    <au><snm>Schütte</snm><fnm>S</fnm></au>
    <au><snm>Bolon</snm><fnm>I</fnm></au>
    <au><snm>Keiser</snm><fnm>O</fnm></au>
    <au><snm>Escher</snm><fnm>G</fnm></au>
    <au><snm>Flahault</snm><fnm>A</fnm></au>
  </aug>
  <source>Journal of Global Health</source>
  <pubdate>2019</pubdate>
  <volume>9</volume>
</bibl>

<bibl id="B7">
  <title><p>Digital Technology and Mobile Applications Impact on Zika and Ebola
  Epidemics Data Sharing and Emergency Response</p></title>
  <aug>
    <au><snm>Tambo</snm><fnm>E</fnm></au>
    <au><snm>Adama</snm><fnm>K</fnm></au>
    <au><snm>Talla</snm><fnm>M</fnm></au>
    <au><snm>CF</snm><fnm>C</fnm></au>
    <au><snm>C.</snm><fnm>F</fnm></au>
  </aug>
  <source>J Health Med Informatics</source>
  <pubdate>2017</pubdate>
  <volume>8</volume>
  <fpage>254</fpage>
</bibl>

<bibl id="B8">
  <title><p>Using NASA Satellite Data to Predict Malaria Outbreaks</p></title>
  <aug>
    <au><snm>Reiny</snm><fnm>S</fnm></au>
  </aug>
  <pubdate>2017</pubdate>
  <url>https://www.nasa.gov/feature/goddard/2017/using-nasa-satellite-data-to-predict-malaria-outbreaks</url>
</bibl>

<bibl id="B9">
  <title><p>Digital Solutions for Malaria Elimination Community of
  Practice</p></title>
  <pubdate>2017</pubdate>
  <url>http://dsme.community/</url>
</bibl>

<bibl id="B10">
  <title><p>Example of layout and flow of individuals being
  screened</p></title>
  <pubdate>2020</pubdate>
  <url>https://www.cdc.gov/coronavirus/2019-ncov/images/hcp/figure-layout-flow-screening.png</url>
</bibl>

<bibl id="B11">
  <title><p>Performing Broad-Based Testing for SARS-CoV-2 in Congregate
  Settings</p></title>
  <pubdate>2020</pubdate>
  <url>https://www.cdc.gov/coronavirus/2019-ncov/hcp/broad-based-testing.html</url>
</bibl>

<bibl id="B12">
  <title><p>NHS Testing and tracing for coronavirus</p></title>
  <pubdate>2020</pubdate>
  <url>https://www.nhs.uk/conditions/coronavirus-covid-19/testing-and-tracing/</url>
</bibl>

<bibl id="B13">
  <title><p>MIT Medical FAQ Testing for COVID-19</p></title>
  <pubdate>2020</pubdate>
  <url>https://medical.mit.edu/faqs/faq-testing-covid-19</url>
</bibl>

<bibl id="B14">
  <title><p>Minnesota Community Testing</p></title>
  <pubdate>2020</pubdate>
  <url>https://www.health.state.mn.us/diseases/coronavirus/testsites/community.html</url>
</bibl>

<bibl id="B15">
  <title><p>Henry Ford COVID-19 Symptoms Testing</p></title>
  <pubdate>2020</pubdate>
  <url>https://www.henryford.com/coronavirus/covid19-symptoms-testing-treatment</url>
</bibl>

<bibl id="B16">
  <title><p>CareWell COVID-19 Testing</p></title>
  <pubdate>2020</pubdate>
  <url>https://www.carewellurgentcare.com/covid-19-response-at-carewell/</url>
</bibl>

<bibl id="B17">
  <title><p>The scientific and ethical feasibility of immunity
  passports</p></title>
  <aug>
    <au><snm>Brown</snm><fnm>B</fnm></au>
    <au><snm>Kelly</snm><fnm>D</fnm></au>
    <au><snm>Wilkinson</snm><fnm>D</fnm></au>
    <au><snm>Savulescu</snm><fnm>J</fnm></au>
  </aug>
  <source>The Lancet Infectious Diseases</source>
  <pubdate>2020</pubdate>
</bibl>

<bibl id="B18">
  <title><p>The Dangerous History of Immunoprivilege</p></title>
  <aug>
    <au><snm>Olivarius</snm><fnm>K</fnm></au>
  </aug>
  <source>NY Times</source>
  <pubdate>2020</pubdate>
  <url>https://www.nytimes.com/2020/04/12/opinion/coronavirus-immunity-passports.html#click=https://t.co/QcXDROj5IL</url>
</bibl>

<bibl id="B19">
  <title><p>Exit through the App Store?</p></title>
  <aug>
    <au><snm>Kind</snm><fnm>C</fnm></au>
  </aug>
  <source>Patterns</source>
  <pubdate>2020</pubdate>
  <volume>1</volume>
  <issue>3</issue>
  <fpage>100054</fpage>
  <url>http://www.sciencedirect.com/science/article/pii/S2666389920300659</url>
</bibl>

<bibl id="B20">
  <title><p>Immunity Certificates: If We Must Have Them, We Must Do It
  Right</p></title>
  <aug>
    <au><snm>Gruener</snm><fnm>D</fnm></au>
  </aug>
  <pubdate>2020</pubdate>
  <url>https://ethics.harvard.edu/files/center-for-ethics/files/12immunitycertificates.pdf</url>
</bibl>

<bibl id="B21">
  <title><p>Health Passport Europe</p></title>
  <pubdate>2020</pubdate>
  <url>https://www.healthpassportireland.ie/</url>
</bibl>

<bibl id="B22">
  <title><p>In Coronavirus Fight, China Gives Citizens a Color Code, With Red
  Flags</p></title>
  <aug>
    <au><snm>Paul Mozur</snm><fnm>RZ</fnm></au>
    <au><snm>Krolik</snm><fnm>A</fnm></au>
  </aug>
  <pubdate>2020</pubdate>
  <url>https://www.nytimes.com/2020/03/01/business/china-coronavirus-surveillance.html</url>
</bibl>

<bibl id="B23">
  <title><p>WHO doesn’t recommend coronavirus passports, because immunity
  remains questionable</p></title>
  <aug>
    <au><snm>Jr.</snm><fnm>BL</fnm></au>
  </aug>
  <pubdate>2020</pubdate>
  <url>https://www.cnbc.com/2020/09/16/who-doesnt-recommend-coronavirus-passports-because-immunity-remains-questionable.html</url>
</bibl>

<bibl id="B24">
  <title><p>"Immunity passports" in the context of COVID-19</p></title>
  <aug>
    <au><cnm>WHO</cnm></au>
  </aug>
  <pubdate>2020</pubdate>
  <url>https://www.who.int/publications/i/item/immunity-passports-in-the-context-of-covid-19</url>
</bibl>

<bibl id="B25">
  <title><p>IBM Digital Health Pass</p></title>
  <pubdate>2020</pubdate>
  <url>https://www.ibm.com/products/digital-health-pass</url>
</bibl>

<bibl id="B26">
  <title><p>V-Health: The People's Passport</p></title>
  <pubdate>2020</pubdate>
  <url>https://v-healthpassport.co.uk/</url>
</bibl>

<bibl id="B27">
  <title><p>WISeKey’s WIShelter Covid-19 Platform Using Digital IDs and
  Blockchain to Help Tourist Destinations Certify Travelers are not
  Infected</p></title>
  <pubdate>2020</pubdate>
  <url>https://www.wisekey.com/press/wisekeys-wishelter-covid-19-platform-using-digital-ids-and-blockchain-to-help-tourist-destin\\ations-certify-travelers-are-not-infected/</url>
</bibl>

<bibl id="B28">
  <title><p>COVI-Pass by Tento Health</p></title>
  <pubdate>2020</pubdate>
  <url>https://tentohealth.com/</url>
</bibl>

<bibl id="B29">
  <title><p>Digital ID immunity passports gain steam around the
  globe</p></title>
  <aug>
    <au><cnm>SecureIdNews</cnm></au>
  </aug>
  <pubdate>2020</pubdate>
  <url>https://www.secureidnews.com/news-item/digital-id-immunity-passports-gain-steam-around-the-globe/</url>
</bibl>

<bibl id="B30">
  <title><p>International monitor: public health identity systems, Ada Lovelace
  Institute</p></title>
  <pubdate>2020</pubdate>
  <url>https://www.adalovelaceinstitute.org/project/international-monitor-public-health-identity-systems/</url>
</bibl>

<bibl id="B31">
  <title><p>Irish-based ROQU Group launches world-first 'Health Passport'
  digital platform to support increased global COVID-19 testing</p></title>
  <aug>
    <au><cnm>PRNewswire</cnm></au>
  </aug>
  <pubdate>2020</pubdate>
  <url>https://www.prnewswire.com/news-releases/irish-based-roqu-group-launches-world-first-health-passport-digital-platform-to-sup\\port-increased-global-covid-19-testing-301120037.html</url>
</bibl>

<bibl id="B32">
  <title><p>ALHOSN: Protecting yourself protects your community</p></title>
  <pubdate>2020</pubdate>
  <url>https://www.alhosnapp.ae/en/home/</url>
</bibl>

<bibl id="B33">
  <title><p>WHO Digital COVID-19 Vaccination Infrastructure</p></title>
  <pubdate>2020</pubdate>
  <url>https://guardtime.com/blog/who-digital-covid-19-vaccination-infrastructure</url>
</bibl>

<bibl id="B34">
  <title><p>COVID-19 Credentials Initiative</p></title>
  <pubdate>2020</pubdate>
  <url>https://www.covidcreds.com/</url>
</bibl>

<bibl id="B35">
  <title><p>Digital health passports roll out for air travelers tracking COVID
  status, vaccination proof</p></title>
  <aug>
    <au><snm>SMITS</snm><fnm>JEANNE</fnm></au>
  </aug>
  <pubdate>2020</pubdate>
  <url>https://www.lifesitenews.com/blogs/digital-health-passports-roll-out-for-air-travelers-tracking-covid-status-vaccination-proof</url>
</bibl>

<bibl id="B36">
  <title><p>CommonPass by The Commons Project</p></title>
  <pubdate>2020</pubdate>
  <url>https://thecommonsproject.org/commonpass</url>
</bibl>

<bibl id="B37">
  <title><p>CommonPass</p></title>
  <pubdate>2020</pubdate>
  <url>https://commonpass.org/</url>
</bibl>

<bibl id="B38">
  <title><p>Health Pass by CLEAR</p></title>
  <pubdate>2020</pubdate>
  <url>https://www.clearme.com/healthpass</url>
</bibl>

<bibl id="B39">
  <title><p>Hi+ Card by TDDS</p></title>
  <pubdate>2020</pubdate>
  <url>https://hicard.travel/</url>
</bibl>

<bibl id="B40">
  <title><p>Clinical Landscape of COVID-19 Testing: Difficult
  Choices</p></title>
  <aug>
    <au><snm>Gandhi</snm><fnm>D</fnm></au>
    <au><snm>Landage</snm><fnm>S</fnm></au>
    <au><snm>Bae</snm><fnm>J</fnm></au>
    <au><snm>Shankar</snm><fnm>S</fnm></au>
    <au><snm>Sukumaran</snm><fnm>R</fnm></au>
    <au><snm>Patwa</snm><fnm>P</fnm></au>
    <au><snm>au2</snm><fnm>STV</fnm></au>
    <au><snm>Katiyar</snm><fnm>P</fnm></au>
    <au><snm>Advani</snm><fnm>S</fnm></au>
    <au><snm>Iyer</snm><fnm>R</fnm></au>
    <au><snm>Anand</snm><fnm>S</fnm></au>
    <au><snm>Mahindra</snm><fnm>A</fnm></au>
    <au><snm>Barbar</snm><fnm>R</fnm></au>
    <au><snm>Singh</snm><fnm>A</fnm></au>
    <au><snm>Raskar</snm><fnm>R</fnm></au>
  </aug>
  <pubdate>2020</pubdate>
</bibl>

<bibl id="B41">
  <title><p>Clinical Laboratory Improvement Amendments (CLIA)</p></title>
  <pubdate>2020</pubdate>
  <url>https://www.cdc.gov/clia/index.html</url>
</bibl>

<bibl id="B42">
  <title><p>Quest Diagnostics, COVID-19</p></title>
  <pubdate>2020</pubdate>
  <url>https://www.questdiagnostics.com/home/Covid-19/</url>
</bibl>

<bibl id="B43">
  <title><p>COVID-19 Laboratory Testing: Medpace Central Labs</p></title>
  <pubdate>2020</pubdate>
  <url>https://www.medpace.com/covid-19-laboratory-testing/</url>
</bibl>

<bibl id="B44">
  <title><p>Flexible, agile and proactive laboratory services dedicated to
  global clinical development</p></title>
  <pubdate>2020</pubdate>
  <url>https://www.iconplc.com/services/laboratories/</url>
</bibl>

<bibl id="B45">
  <title><p>Syneos Health - Collaborating to Address the Challenge of
  COVID-19</p></title>
  <pubdate>2020</pubdate>
  <url>https://www.syneoshealth.com/covid-19-resource-center</url>
</bibl>

<bibl id="B46">
  <title><p>Verily: Delivering COVID-19 screening and testing through Project
  Baseline and community-based sites</p></title>
  <pubdate>2020</pubdate>
  <url>https://verily.com/solutions/covid-19-testing/</url>
</bibl>

<bibl id="B47">
  <title><p>2019 Novel Coronavirus Disease (COVID-19)</p></title>
  <pubdate>2020</pubdate>
  <url>https://www.bioreference.com/coronavirus/</url>
</bibl>

<bibl id="B48">
  <title><p>Find Diseases &amp Conditions</p></title>
  <pubdate>2020</pubdate>
  <url>https://www.mayoclinic.org</url>
</bibl>

<bibl id="B49">
  <title><p>Amherst College COVID-19 Dashboard</p></title>
  <pubdate>2020</pubdate>
  <url>https://www.amherst.edu/news/covid-19/dashboard</url>
</bibl>

<bibl id="B50">
  <title><p>We Rate COVID-19 Dashboards</p></title>
  <pubdate>2020</pubdate>
  <url>https://www.ratecoviddashboard.com/</url>
</bibl>

<bibl id="B51">
  <title><p>Wagner College COVID-19 Dashboard</p></title>
  <pubdate>2020</pubdate>
  <url>https://sites.google.com/wagner.edu/covid-19casetracker</url>
</bibl>

<bibl id="B52">
  <title><p>Tulane University COVID-19 Dashboard</p></title>
  <pubdate>2020</pubdate>
  <url>https://tulane.edu/covid-19/dashboard</url>
</bibl>

<bibl id="B53">
  <title><p>Ohio State University COVID-19 Dashboard</p></title>
  <pubdate>2020</pubdate>
  <url>https://news.osu.edu/a-deeper-dive-into-ohio-states-top-rated-covid-19-testing-data-dashboard/</url>
</bibl>

<bibl id="B54">
  <title><p>George Mason University COVID-19 Dashboard</p></title>
  <pubdate>2020</pubdate>
  <url>https://www2.gmu.edu/campus-covid-19-data</url>
</bibl>

<bibl id="B55">
  <title><p>CDC COVID Data Tracker</p></title>
  <pubdate>2020</pubdate>
  <url>https://covid.cdc.gov/covid-data-tracker/?CDC_AA_refVal=https%3A%2F%2Fwww.cdc.gov%2Fcoronavirus%2F2019-ncov%2Fcases-updates%2Fcases-in-us.html#cases_casesper100klast7days</url>
</bibl>

<bibl id="B56">
  <title><p>Carnegie Mellon University COVID-19 Dashboard</p></title>
  <pubdate>2020</pubdate>
  <url>https://covidcast.cmu.edu/?sensor=doctor-visits-smoothed_adj_cli&level=county&date=20201027&signalType=value&encoding=color&mode=single&region=42003</url>
</bibl>

<bibl id="B57">
  <title><p>Coronavirus disease (COVID-19): Outbreak update, Government of
  Canada</p></title>
  <pubdate>2020</pubdate>
  <url>https://www.canada.ca/en/public-health/services/diseases/2019-novel-coronavirus-infection.html?topic=tilelink</url>
</bibl>

<bibl id="B58">
  <title><p>Evolution of COVID-19, CBC Radio-Canada</p></title>
  <pubdate>2020</pubdate>
  <url>https://ici.radio-canada.ca/info/2020/coronavirus-covid-19-pandemie-cas-carte-maladie-symptomes-propagation/index-en.html</url>
</bibl>

<bibl id="B59">
  <title><p>COVID-19 State wise Status, Ministry of Health and Family Welfare,
  India</p></title>
  <pubdate>2020</pubdate>
  <url>https://www.mohfw.gov.in/</url>
</bibl>

<bibl id="B60">
  <title><p>COVID-19 : Current cases, New Zealand Ministry of
  Health</p></title>
  <pubdate>2020</pubdate>
  <url>https://www.health.govt.nz/our-work/diseases-and-conditions/covid-19-novel-coronavirus/covid-19-data-and-statistics/covid-19-current-cases</url>
</bibl>

<bibl id="B61">
  <title><p>France COVID-19 Dashboard</p></title>
  <pubdate>2020</pubdate>
  <url>https://dashboard.covid19.data.gouv.fr/vue-d-ensemble?location=FRA</url>
</bibl>

<bibl id="B62">
  <title><p>Worldometer</p></title>
  <pubdate>2020</pubdate>
  <url>https://www.worldometers.info/coronavirus/</url>
</bibl>

<bibl id="B63">
  <title><p>thebaselab</p></title>
  <pubdate>2020</pubdate>
  <url>https://coronavirus.thebaselab.com</url>
</bibl>

<bibl id="B64">
  <title><p>CDC COVID-19 Testing Overview</p></title>
  <pubdate>2020</pubdate>
  <url>https://www.cdc.gov/coronavirus/2019-ncov/symptoms-testing/testing.html</url>
</bibl>

<bibl id="B65">
  <title><p>Apple COVID-19 Screening Tool</p></title>
  <pubdate>2020</pubdate>
  <url>https://covid19.apple.com/screening#donotlink</url>
</bibl>

<bibl id="B66">
  <title><p>Castlight Inc. COVID-19 Test Site Finder: Get Information on
  Coronavirus Testing Near You</p></title>
  <pubdate>2020</pubdate>
  <url>https://my.castlighthealth.com/corona-virus-testing-sites/</url>
</bibl>

<bibl id="B67">
  <title><p>Community-Based Testing Sites for COVID-19</p></title>
  <pubdate>2020</pubdate>
  <url>https://www.hhs.gov/coronavirus/community-based-testing-sites/index.html</url>
</bibl>

<bibl id="B68">
  <title><p>Project Baseline by verily</p></title>
  <pubdate>2020</pubdate>
  <url>https://www.projectbaseline.com/study/covid-19/</url>
</bibl>

<bibl id="B69">
  <title><p>Comparing privacy laws: GDPR v. CCPA</p></title>
  <pubdate>2020</pubdate>
  <url>https://fpf.org/wp-content/uploads/2018/11/GDPR_CCPA_Comparison-Guide.pdf</url>
</bibl>

<bibl id="B70">
  <title><p>Deliberation on HIPAA Privacy Rule</p></title>
  <pubdate>2020</pubdate>
  <url>https://www.hhs.gov/hipaa/for-professionals/faq/405/is-this-more-protective-state-law-preempted-by-the-privacy-rule/index.html</url>
</bibl>

<bibl id="B71">
  <title><p>Does the GDPR apply to companies outside of the EU?</p></title>
  <pubdate>2020</pubdate>
  <url>https://gdpr.eu/companies-outside-of-europe</url>
</bibl>

<bibl id="B72">
  <title><p>Healthcare Data Breach Statistics, HIPAA Journal</p></title>
  <pubdate>2020</pubdate>
  <url>https://www.hipaajournal.com/healthcare-data-breach-statistics/</url>
</bibl>

<bibl id="B73">
  <title><p>One in Four US Consumers Have Had Their Healthcare Data Breached,
  Accenture Survey Reveals</p></title>
  <pubdate>2020</pubdate>
  <url>https://newsroom.accenture.com/subjects/technology/one-in-four-us-consumers-have-had-their-healthcare-data-breached-accenture-survey-reveals.html</url>
</bibl>

<bibl id="B74">
  <title><p>Top 10 Biggest Healthcare Data Breaches of All Time</p></title>
  <pubdate>2020</pubdate>
  <url>https://digitalguardian.com/blog/top-10-biggest-healthcare-data-breaches-all-time</url>
</bibl>

<bibl id="B75">
  <title><p>John Hopkins University Mortality Analyses</p></title>
  <pubdate>2020</pubdate>
  <url>https://coronavirus.jhu.edu/data/mortality</url>
</bibl>

<bibl id="B76">
  <title><p>COVID-19: Lockdown across India, in line with WHO
  guidance</p></title>
  <pubdate>2020</pubdate>
  <url>https://news.un.org/en/story/2020/03/1060132</url>
</bibl>

<bibl id="B77">
  <title><p>COVID-19 in India: Lockdown, Legal Challenges, and Disparate
  Impacts</p></title>
  <aug>
    <au><snm>Grover</snm><fnm>A</fnm></au>
  </aug>
  <pubdate>2020</pubdate>
  <url>https://blog.petrieflom.law.harvard.edu/2020/05/18/india-global-responses-covid19/</url>
</bibl>

<bibl id="B78">
  <title><p>THE CONSTITUTION OF INDIA</p></title>
  <pubdate>2020</pubdate>
  <url>https://www.india.gov.in/sites/upload_files/npi/files/coi_part_full.pdf</url>
</bibl>

<bibl id="B79">
  <title><p>Frequently Asked Questions on Aarogya Setu App</p></title>
  <pubdate>2020</pubdate>
  <url>https://covid19.assam.gov.in/wp-content/uploads/2020/05/Aarogya_Setu_FAQ.pdf</url>
</bibl>

<bibl id="B80">
  <title><p>DATA PROTECTION LAWS OF THE WORLD</p></title>
  <pubdate>2020</pubdate>
  <url>https://www.dlapiperdataprotection.com/index.html?t=law&c=VN</url>
</bibl>

<bibl id="B81">
  <title><p>Data Privacy &amp Security Survey</p></title>
  <aug>
    <au><snm>McKenzie</snm><fnm>B</fnm></au>
  </aug>
  <pubdate>2020</pubdate>
  <url>https://www.bakermckenzie.com/-/media/files/insight/publications/2020/04/covid19-data-privacy--security-survey17-april.pdf</url>
</bibl>

<bibl id="B82">
  <title><p>"Protect yourself" Ministry of Health Vietnam</p></title>
  <pubdate>2020</pubdate>
  <url>https://bluezone.ai</url>
</bibl>

<bibl id="B83">
  <title><p>Pervasive personal data collection at the heart of South Korea’s
  COVID-19 success may not translate</p></title>
  <aug>
    <au><snm>Helen Chan</snm><fnm>RIE</fnm></au>
  </aug>
  <pubdate>2020</pubdate>
  <url>https://blogs.thomsonreuters.com/answerson/south-korea-covid-19-data-privacy/</url>
</bibl>

<bibl id="B84">
  <title><p>Privacy vs. Pandemic Control in South Korea</p></title>
  <aug>
    <au><snm>Panakal</snm><fnm>DD</fnm></au>
  </aug>
  <pubdate>2020</pubdate>
  <url>https://www.natlawreview.com/article/privacy-vs-pandemic-control-south-korea</url>
</bibl>

<bibl id="B85">
  <title><p>Coronavirus tracing app a test for privacy-minded
  Germany</p></title>
  <aug>
    <au><snm>Frank Jordans</snm><fnm>AP</fnm></au>
  </aug>
  <pubdate>2020</pubdate>
  <url>https://abcnews.go.com/Health/wireStory/coronavirus-tracing-app-test-privacy-minded-germany-71270334</url>
</bibl>

<bibl id="B86">
  <title><p>Mobile applications in support of contact tracing for COVID-19,
  European Centre for Disease Prevention and Control</p></title>
  <pubdate>2020</pubdate>
  <url>https://www.ecdc.europa.eu/sites/default/files/documents/covid-19-mobile-applications-contact-tracing.pdf</url>
</bibl>

<bibl id="B87">
  <title><p>Emerging COVID-19 success story: Germany’s strong enabling
  environment</p></title>
  <aug>
    <au><snm>Lothar Wieler</snm><fnm>RG</fnm></au>
  </aug>
  <pubdate>2020</pubdate>
  <url>https://ourworldindata.org/covid-exemplar-germany</url>
</bibl>

<bibl id="B88">
  <title><p>Navigating China Episode 14: New draft national, harmonised data
  protection law for Mainland China Navigating China: The digital
  journey</p></title>
  <aug>
    <au><snm>Carolyn Bigg</snm><fnm>FS</fnm></au>
  </aug>
  <pubdate>2020</pubdate>
  <url>https://www.dlapiper.com/en/us/insights/publications/2020/10/navigating-china-episode-14/</url>
</bibl>

<bibl id="B89">
  <title><p>What is a Data Silo and Why is It Bad for Your
  Organization?</p></title>
  <aug>
    <au><cnm>Alienor</cnm></au>
  </aug>
  <pubdate>2020</pubdate>
  <url>https://www.plixer.com/blog/data-silo-what-is-it-why-is-it-bad/</url>
</bibl>

<bibl id="B90">
  <title><p>What are Data Silos?</p></title>
  <pubdate>2020</pubdate>
  <url>https://www.talend.com/resources/what-are-data-silos/</url>
</bibl>

<bibl id="B91">
  <title><p>Operationalizing Data Lake Privacy Governance for Value
  Creation</p></title>
  <aug>
    <au><snm>Turajski</snm><fnm>N</fnm></au>
  </aug>
  <pubdate>2020</pubdate>
  <url>https://www.talend.com/resources/what-are-data-silos/</url>
</bibl>

<bibl id="B92">
  <title><p>No Peek: A Survey of private distributed deep learning</p></title>
  <aug>
    <au><snm>Vepakomma</snm><fnm>P</fnm></au>
    <au><snm>Swedish</snm><fnm>T</fnm></au>
    <au><snm>Raskar</snm><fnm>R</fnm></au>
    <au><snm>Gupta</snm><fnm>O</fnm></au>
    <au><snm>Dubey</snm><fnm>A</fnm></au>
  </aug>
  <pubdate>2018</pubdate>
</bibl>

<bibl id="B93">
  <title><p>Artificial Intelligence in the Battle against Coronavirus
  (COVID-19): A Survey and Future Research Directions</p></title>
  <aug>
    <au><snm>Nguyen</snm><fnm>TT</fnm></au>
  </aug>
  <pubdate>2020</pubdate>
  <url>https://www.researchgate.net/publication/340487417_Artificial_Intelligence_in_the_Battle_against_Coronavirus_COVID-19_A_Survey_and_Future_Research_Directions</url>
</bibl>

<bibl id="B94">
  <title><p>Point-of-care testing for COVID-19 using SHERLOCK
  diagnostics</p></title>
  <aug>
    <au><snm>Joung</snm><fnm>J</fnm></au>
    <au><snm>Ladha</snm><fnm>A</fnm></au>
    <au><snm>Saito</snm><fnm>M</fnm></au>
    <au><snm>Segel</snm><fnm>M</fnm></au>
    <au><snm>Bruneau</snm><fnm>R</fnm></au>
    <au><snm>Huang</snm><fnm>MlW</fnm></au>
    <au><snm>Kim</snm><fnm>NG</fnm></au>
    <au><snm>Yu</snm><fnm>X</fnm></au>
    <au><snm>Li</snm><fnm>J</fnm></au>
    <au><snm>Walker</snm><fnm>BD</fnm></au>
    <au><snm>Greninger</snm><fnm>AL</fnm></au>
    <au><snm>Jerome</snm><fnm>KR</fnm></au>
    <au><snm>Gootenberg</snm><fnm>JS</fnm></au>
    <au><snm>Abudayyeh</snm><fnm>OO</fnm></au>
    <au><snm>Zhang</snm><fnm>F</fnm></au>
  </aug>
  <source>medRxiv</source>
  <publisher>Cold Spring Harbor Laboratory Press</publisher>
  <pubdate>2020</pubdate>
  <url>https://www.medrxiv.org/content/early/2020/05/08/2020.05.04.20091231</url>
</bibl>

<bibl id="B95">
  <title><p>Artificial intelligence model detects asymptomatic Covid-19
  infections through cellphone-recorded coughs</p></title>
  <aug>
    <au><snm>Chu</snm><fnm>J</fnm></au>
  </aug>
  <pubdate>2020</pubdate>
  <url>https://news.mit.edu/2020/covid-19-cough-cellphone-detection-1029</url>
</bibl>

</refgrp>
} % end of \BMCxmlcomment
\bibliographystyle{bmc-mathphys}
% for author-year bibliography (bmc-mathphys or spbasic)
% a) write to bib file (bmc-mathphys only)
% @settings{label, options="nameyear"}

% or include bibliography directly:
% \begin{thebibliography}
% \bibitem{b1}
% \end{thebibliography}

\end{backmatter}
\end{document}